\documentstyle[aps,multicol]{revtex}
\renewcommand{\narrowtext}{\begin{multicols}{2} \global\columnwidth20.5pc}
\renewcommand{\widetext}{\end{multicols} \global\columnwidth42.5pc}

\input{epsf.tex}
\def\inseps#1#2{\def\epsfsize##1##2{#2##1} \centerline{\epsfbox{#1}}}

\def\top#1{\vskip #1\begin{picture}(290,80)(80,500)\thinlines \put(
65,500){\line( 1, 0){255}}\put(320,500){\line( 0, 1){
5}}\end{picture}}
\def\bottom#1{\vskip #1\begin{picture}(290,80)(80,500)\thinlines \put(
330,500){\line( 1, 0){255}}\put(330,500){\line( 0, -1){
5}}\end{picture}}

\begin{document}
\draft

\title{Theory of Photoluminescence of the $\nu=1$ Quantum Hall
State:\\ Excitons, Spin-Waves and Spin-Textures}

\author{N.R. Cooper$^{(1)}$ and D.B. Chklovskii$^{(2)}$} 
\address{(1) Institut Laue-Langevin, Avenue Des Martyrs, 
B.P. 156, 38042 Grenoble, Cedex 9, France. \\(2) Lyman Laboratory of
Physics, Harvard University, Cambridge, MA 02138.}

\date{\today} 
\maketitle

\begin{abstract}

We study the theory of intrinsic photoluminescence of two-dimensional
electron systems in the vicinity of the $\nu=1$ quantum Hall state.
We focus predominantly on the recombination of a band of initial
``excitonic states'' that are the low-lying energy states of our model
at $\nu=1$. It is shown that the recombination of excitonic states can
account for recent observations of the polarization-resolved spectra
of a high-mobility GaAs quantum well.  The asymmetric broadening of
the spectral line in the $\sigma_-$ polarization is explained to be
the result of the ``shake-up'' of spin-waves upon radiative
recombination of excitonic states.  We derive line shapes for the
recombination of excitonic states in the presence of long-range
disorder that compare favourably with the experimental observations.
We also discuss the stabilities and recombination spectra of other
(``charged'') initial states of our model.  An additional high-energy
line observed in experiment is shown to be consistent with the
recombination of a positively-charged state.  The recombination
spectrum of a negatively-charged initial state, predicted by our model
but not observed in the present experiments, is shown to provide a
direct measure of the formation energy of the smallest ``charged
spin-texture'' of the $\nu=1$ state.

\end{abstract}


\narrowtext

\section{Introduction}

\label{sec:introduction}

Continuing improvements in the quality of quantum-well devices are
leading to increasing resolution of the intrinsic photoluminescence
spectra of two-dimensional electron systems in the extreme quantum
regime.  It is now well-established that features in the
photoluminescence spectra are related to the appearance of the integer
and fractional quantum Hall states and the insulating phase associated
with the magnetically-induced Wigner
crystal\cite{clarkreview,heimanb,turberfieldpol,goldbergoverview}. The
possibility of extracting information on the properties of these
strongly-correlated phases from the photoluminescence spectra has
stimulated a great deal of recent experimental and theoretical
interest in this technique.

The interpretation of photoluminescence spectra requires an
understanding of the energy eigenstates of a valence-band hole in the
presence of the electron gas. Due to the strong many-body interactions
that are important in the extreme quantum regime of these systems,
this presents an essentially strongly-coupled many-body problem and
the interpretation of spectral structure is extremely difficult.  The
theories that have been developed to address this issue fall into two
broad categories.  Certain theories treat the inter-particle
correlations approximately through the use of some form of mean-field
description of the interactions\cite{uenoyamasham,bauer,whittaker}.
Such an approach has been shown to successfully account for
oscillations in the mean position of the luminescence line in the
integer quantum Hall regime of disordered samples\cite{uenoyamasham}.
The fractional quantum Hall and Wigner crystal regimes cannot be
described within such a mean-field approach.  To treat these cases,
other theories have been developed which attempt to treat the
inter-particle correlations more
accurately\cite{apalkovpikusefros,apalkovprb,chenquinn95,chenprb94,zangbirman95,macrezkell,tatarinova,rashbaanyon,cooperwc}.
In these theories, a simplified model is usually adopted in which the
electrons and photo-excited hole are restricted to states in the
lowest Landau level.  For the most part, the resulting many-body
problem has been treated by numerical diagonalization of small
systems\cite{apalkovpikusefros,apalkovprb,chenquinn95,chenprb94,zangbirman95,macrezkell},
though some approximate analytic treatments have been proposed in the
fractional quantum Hall\cite{tatarinova,rashbaanyon} and Wigner
crystal\cite{cooperwc} regimes. Despite the great deal of theoretical
effort, the comparison between the theoretical and experimental
photoluminescence spectra of high-mobility samples is still rather
unsatisfactory, with even qualitative features of the observed spectra
still not convincingly accounted for.  (We note that ``acceptor-bound
photoluminescence'' spectra are somewhat better
understood\cite{buhmannfqhe,kuktriangle}. This experimental technique
is quite different from intrinsic photoluminescence which we study
here.)

It is the purpose of this paper to show that the intrinsic
photoluminescence spectra of two-dimensional systems close to the
integer filling fraction $\nu=1$ contain interesting and non-trivial
structure (the filling fraction is defined by $\nu\equiv n h/eB$,
where $n$ is the electron density and $eB/h$ the density of flux
quanta).  From a theoretical point of view, this is a much simpler
filling fraction to study than the fractional quantum Hall and Wigner
crystal regimes, yet still poses a non-trivial problem due to the
importance of strong correlations in determining the low-energy
excitations at this filling fraction.

\begin{figure}
\inseps{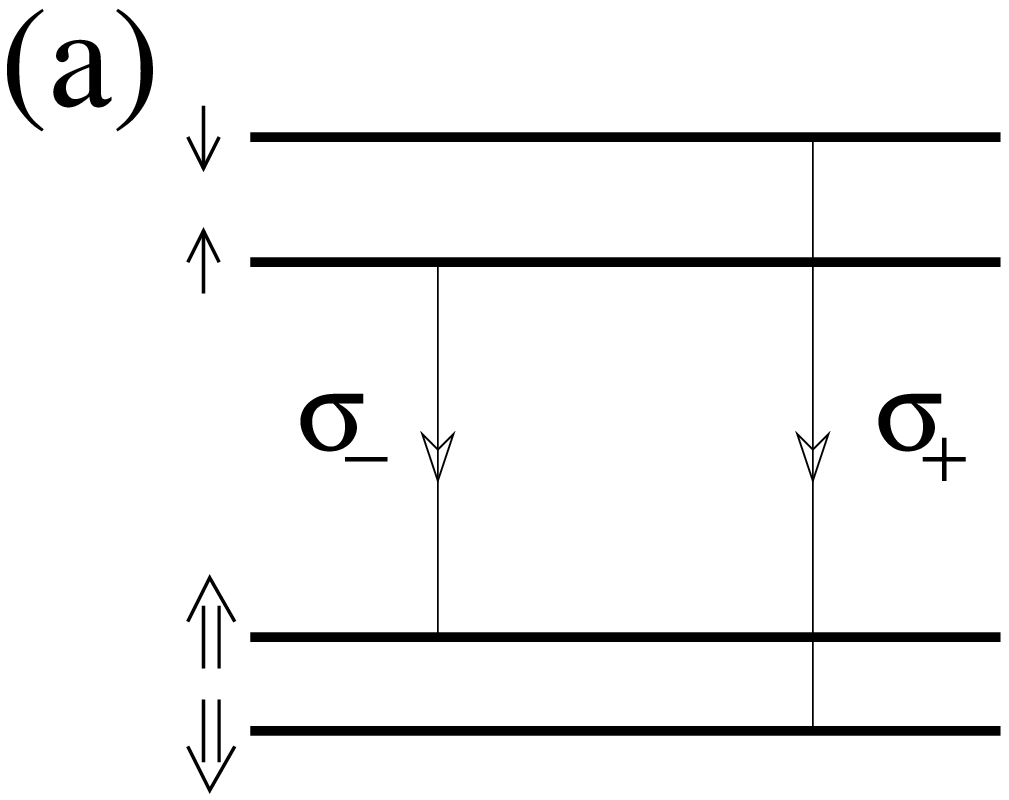}{0.38}
\inseps{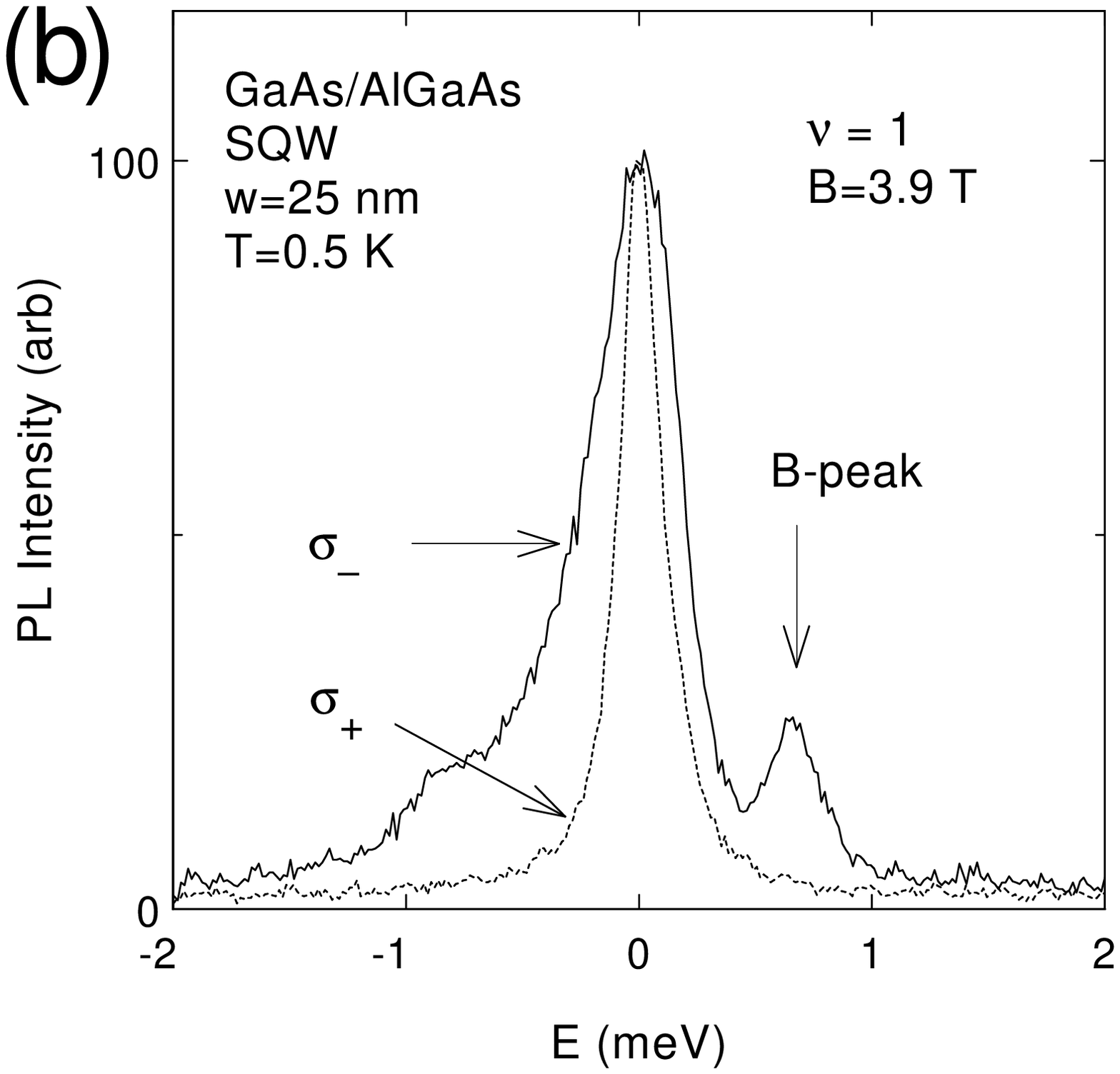}{0.40}
\caption{(a) Schematic diagram of the low-energy
interband transitions in a GaAs quantum well.  The recombination of
electrons of the two spin orientations of the lowest Landau level
gives rise to the two different circular polarizations.  (b)
Low-temperature photoluminescence spectra of a GaAs quantum well at
$\nu=1$, from the experiments reported in
Ref.~\protect\onlinecite{plentz}.  In the $\sigma_+$ circular
polarization, a narrow line is observed with a line shape similar to
that observed away from $\nu=1$.  In the $\sigma_-$ circular
polarization, the spectrum acquires a strong low-energy broadening as
compared to the spectra away from $\nu=1$; an additional high-energy
peak (peak B) appears at low temperatures.  Energies are measured
relative to the energies at which the intensities are maximal.}
\label{fig:experiment}
\end{figure}

In recent photoluminescence experiments on a very high-mobility GaAs
quantum well, extremely narrow line widths have been achieved and very
interesting low-energy structure has been resolved\cite{plentz}. It is
found that as the filling fraction of the sample is swept through
$\nu=1$, the photoluminescence spectrum displays very intriguing
behaviour.  The evolution of the spectrum is quite different in the
two circular polarizations, which originate from the recombination of
a hole with electrons of the two spin-polarizations, as illustrated in
Fig.~\ref{fig:experiment}(a).  In one polarization ($\sigma_+$), no
significant features are observed in the spectrum at $\nu=1$; the
integrated intensity shows a weak minimum, but the line shape is
almost unchanged.  In the other polarization ($\sigma_-$), a much more
dramatic evolution is observed: the integrated intensity also
decreases slightly, but, at the same time, the main spectral line
becomes strongly broadened on the low-energy side; at the lowest
temperatures, an additional high-energy peak appears.  Figure~1(b)
shows the spectra observed in the experiments reported in
Ref.~\onlinecite{plentz} at the filling fraction $\nu=1$ for both
circular polarizations; the additional high-energy peak is labelled
the ``B-peak''.

These observations cannot be accounted for within the existing
``mean-field'' theories of photoluminescence in the integer quantum
Hall regime\cite{uenoyamasham,bauer,whittaker}. In the first place,
these theories treat the spin-degree of freedom of the electrons in
such a way that no polarization-dependent effects can appear.
Moreover, strong correlations are likely to be important in
determining the structure observed in photoluminescence at $\nu=1$,
since it is now clear that the properties of typical GaAs systems at
this filling fraction are dominated by interactions, the single
particle gap (the bare electron Zeeman energy) being very much smaller
than the interaction energy scale.  In fact, as suggested in
Ref.\onlinecite{sondhi}, the $\nu=1$ state is better viewed as a
strongly-correlated state similar to the incompressible states at
fractional filling fraction, since, even for a vanishing
single-particle gap, a quantized Hall effect would appear at $\nu=1$
as a result of the electron-electron repulsion.  This state has
received a great deal of recent
theoretical\cite{sondhi,fertigskyrmion,macdonaldskyrmion} and
experimental interest\cite{barrett,schmeller,aifer}, in an effort to
understand the effects of the electron spin on the properties of the
low-energy excitations.  In this light, it is of particular interest
to understand the origin of the structure appearing in the above
photoluminescence spectra, since this technique separately probes the
two spin states of the electrons.

Motivated by these experiments, we study the theory of the intrinsic
photoluminescence of two-dimensional electron systems at filling
fractions close to $\nu=1$.  We will follow the models used in the
fractional quantum Hall and Wigner crystal regimes and neglect Landau
level mixing for the electrons, but will take full account of the
inter-particle correlations.  The model that one obtains within this
approximation is much simpler to analyse at $\nu=1$ than in either of
these other two regimes. Therefore, at the very least, the study of
this model at $\nu=1$ is the most natural way in which to test the
applicability of the underlying assumptions of the theories for
photoluminescence in the fractional quantum Hall and Wigner crystal
regimes.  Moreover, as we shall see, the photoluminescence spectrum at
$\nu=1$ retains non-trivial structure related to the low-energy
excitations of this state and is therefore of great interest in
itself.  A similar approach to photoluminescence at $\nu=1$ has been
discussed in Ref.\onlinecite{muzykantskii}.  This work did not address
the polarization-dependence of the photoluminescence spectrum.  We
study these issues in some detail, and compare our predictions with
the experimental observations described above.  We show that one can
account for all of the qualitative features observed in the
experiment.

The outline of the paper is as follows.  In Sec.~\ref{sec:description}
we motivate the model that we will study, and discuss its relationship
with other models of photoluminescence in the extreme quantum regime.
In Sec.~\ref{sec:exciton}, we study the predictions of this model at a
filling fraction of exactly $\nu=1$.  We argue that for the sample
studied in Ref.~\cite{plentz}, and for all samples in which the
valence-band hole is close to the electron gas compared to the typical
electron-electron spacing, the most important initial states are
``excitonic states'' (as we choose to name them).  These are states in
which the Landau level of spin-$\uparrow$ electrons is fully occupied,
and the valence-band hole binds with a spin-$\downarrow$ electron to
form an exciton.  In the remainder of this section we study the
photoluminescence spectrum arising from the recombination of these
excitonic states.  This is the main part of the paper, and contains
our most important conclusions with regard to the experimental
observations.  The radiative recombination of the excitonic states is
shown to be quite different in the two circular polarizations.  In the
$\sigma_+$ polarization, the hole recombines with the
spin-$\downarrow$ electron to which it is bound, leaving an
undisturbed Landau level of spin-$\uparrow$ electrons.  In the
$\sigma_-$ polarization, the hole recombines with one of the
spin-$\uparrow$ electrons, and a single spin-reversal is left in the
final state.  The photoluminescence spectrum in this polarization
becomes broadened to low energy due to the ``shake-up'' of these
spin-waves.  We argue that the polarization-dependence of the main
recombination line in the spectra of Fig.~\ref{fig:experiment}(b) can
be accounted for in terms of the recombination of excitonic states:
the recombination line in the $\sigma_-$ polarization is broadened to
low energy due to the shake-up of spin-waves, while the line in the
$\sigma_+$ polarization remains narrow (with a width limited only by
disorder).  We derive the line shapes for a disorder-free system as a
function of the separation between valence-band hole and electron gas
and the extent of Landau level mixing for the hole.  We show that
disorder arising from the remote ionized donor impurities is likely to
have an important effect on the width of this line, and derive line
shapes for the recombination in the two polarizations taking account
of this disorder.  The line shapes compare favourably with the
experimental observations shown in Fig.~\ref{fig:experiment}(b).

In Sec.~\ref{sec:charges}, we turn our attention to quite different
initial states, in which the valence-band hole forms a
positively-charged or negatively-charged complex.  These states can
have lower energies than the excitonic states if the filling fraction
is slightly less than or greater than $\nu=1$ (when some
quasiparticles are present), and can then be important for
photoluminescence.  We show that the high-energy line (peak B) in
Fig.~\ref{fig:experiment}(b) is consistent with the recombination of a
positively-charged initial state in which there are no
spin-$\downarrow$ electrons in the vicinity of the hole.  In this
case, our calculations include corrections arising from Landau level
mixing for the {\it electrons}. These are shown to change the position
of this recombination line relative to that of the excitonic states.
The recombination spectrum of a negatively-charged initial state is
shown, from numerical studies, to contain structure that measures the
formation energy of the smallest ``charged
spin-texture''\cite{sondhi,fertigskyrmion,macdonaldskyrmion} of the
$\nu=1$ state.  There is no clear evidence for this initial state in
the present experiments.  We discuss the type of sample and the
conditions under which this initial state might be more stable and its
recombination could be observed.  Finally, Sec.~\ref{sec:summary}
contains a summary of the main points of the paper.

\section{Description of the model}
\label{sec:description}

We aim to develop a theory that can account for the photoluminescence
of high-mobility quantum wells in the vicinity of $\nu=1$.  In the
experiments of Ref.~\onlinecite{plentz}, and in experiments on similar
GaAs quantum wells\cite{goldbergoverview}, recombination is observed
between the two lowest-energy electron states (the two
spin-polarizations of the lowest Landau level of the lowest subband
state), and the two lowest-energy hole states.  These two hole states
originate from the heavy-hole states of the valence band, but are
strongly mixed with the light-hole states due to the quantum well
confinement\cite{ekenbergaltarelli}. Typically, sufficiently low
excitation powers are used that the density of holes is extremely
small (compared to the density of electrons) and they may be
considered to be independent.

To represent these systems, we will study a model in which the
electrons are confined to a single subband and carry a spin of $1/2$,
and there is a single hole, which may be in one of the two states
($\Uparrow$ or $\Downarrow$)\cite{footnote}. Since we consider the
recombination of a single photo-excited hole, the ``spin'' label of
the hole will play no role other than to define the polarization in
which the hole can recombine [see Fig.~\ref{fig:experiment}(a)].  We
will therefore ignore this label, and leave it to be understood that
when we discuss a recombination process with a spin-$\uparrow$
(spin-$\downarrow$) electron the hole must be in the spin-$\Uparrow$
(spin-$\Downarrow$) state.

For the most part, we will assume that following photo-excitation the
system is able to relax to thermal equilibrium before the hole
recombines.  In this case, one can understand the photoluminescence
spectrum by identifying the low-lying energy eigenstates of a single
photo-excited hole in the presence of the electron gas, and studying
the processes by which these states can decay radiatively.  The rate
of each interband transition is determined by the matrix element of
the electric dipole operator between the initial and final states.
Within the effective mass approximation, this is proportional to the
matrix element of one of the operators
\begin{eqnarray} 
\label{eq:lminus}
L_- & \equiv & \int d^2\bbox{r}\; \psi_{e\uparrow}(\bbox{r}) \psi_{h \Uparrow}
(\bbox{r}), \\ 
\label{eq:lplus} L_+ & \equiv & \int d^2\bbox{r} \;
\psi_{e\downarrow}(\bbox{r}) \psi_{h\Downarrow} (\bbox{r}),
\end{eqnarray}
between the in-plane envelope functions of the initial and final
states, $\psi_{(e,h)\sigma}(\bbox{r})$ being the electron and hole
field annihilation operators.  The absolute transition rate depends on
the overlap of the electron and hole subband wavefunctions, and on the
form of the electron and hole wavefunctions on an atomic scale, which
may differ for the two polarizations.  We will study only the
contributions arising from the matrix elements of $L_\pm$, which are
sufficient to determine the line shapes in the two polarizations.

Due to the importance of many-body interactions in the quantized Hall
regime of these systems, to make progress one must restrict attention
to a somewhat simplified model of the initial and final states of the
photoluminescence process.  A natural model to study which retains the
effects of many-body correlations is one in which the electrons and
the hole are restricted to the lowest Landau level and move in a
single plane.  Such a model is motivated by the success of similar
approximations in accounting for the qualitative and quantitative
properties of the $\nu=1$ state and the incompressible states at
fractional filling fractions\cite{prangeandgirvin}.  However, various
authors\cite{bychkovrashba,paquetriceueda,dzyubenkolozovik,macrezkell,apalkovprb}
have shown that for {\it spin-polarized} electrons and holes
restricted to a single Landau level and with the same quantum well
envelope functions, a ``hidden symmetry'' leads to the result that
photoluminescence contains no spectroscopic structure: the spectrum
consists of a single line at an energy that is independent of the
state, or even the presence, of the electron gas.  In the present
case, the electrons are not spin-polarized.  However, it is
straightforward to show that a similar symmetry applies for an
arbitrary number of spin components for electrons and holes, provided
the interactions conserve the spin of each particle: the spectrum
consists of a series of sharp lines, at energies which are independent
of the state or presence of the electron gas (these are therefore the
energies of each allowed interband transition for an empty quantum
well).

In order to obtain a non-trivial photoluminescence spectrum, it is
essential to study a model that breaks this symmetry.  There are two
clear mechanisms by which this occurs in practice.  Firstly, through
Landau level coupling for the electrons or hole; in GaAs quantum
wells, this is likely to be more important for the hole than for the
electrons due to the much smaller cyclotron energy of the valence band
compared to that of the conduction band.  Secondly, due to the
asymmetry of the single-side-doped quantum wells and single
heterojunctions used in the experiments, the electrons and holes do
not move in the same plane; the hole is pulled somewhat away from the
electron layer\cite{vietbirman}. Note that the presence of a
disordered potential does not break the symmetry, and one of the above
two mechanisms must be introduced.

It is common in theories of photoluminescence in the fractional
quantum Hall regime to retain Landau quantization for both electrons
and holes and to break the ``hidden symmetry'' by introducing a
separation $d$ between the planes in which the electrons and hole
move\cite{apalkovprb,chenprb94}. In our work, we will restrict the
electrons to states in the lowest Landau level, and will also assume
that the electrons and the hole are confined to planes that are
separated by a distance $d$.  We will not, however, impose the
restriction that the hole is in the lowest Landau level. We take
account of Landau level mixing for the valence-band hole by assuming
its in-plane dispersion to be parabolic with an effective mass $m_h$.
Thus we retain two mechanisms by which the hidden symmetry is broken.
We will discuss how the photoluminescence spectrum depends on the
parameters, $d$ and $m_h$.  For quantitative comparisons of our theory
with the experimental observations reproduced in
Fig.~\ref{fig:experiment}(b), we will choose $d$ to be the separation
between the centres of the electron and hole subband wavefunctions in
the quantum well used in these experiments, which is approximately
$60\mbox{\AA}$\cite{flavio} and is therefore small compared to the
magnetic length $\ell=130\mbox{\AA}$ under these conditions (the
magnetic length, $\ell\equiv \sqrt{\hbar/eB}$, is a measure of the
size of a single-particle state in the lowest Landau level and is
therefore a fundamental lengthscale in our model). In the absence of
detailed knowledge of the valence band dispersion, which depends
strongly on the shape of the quantum well\cite{eisensteinholemass}, we
will choose the value $m_h=0.34m_0$ for both hole states; this is
typical of the masses measured in experiment\cite{eisensteinholemass}
and is the value used in theoretical studies of related
problems\cite{wojshawrylak}. Thus, under these conditions the ratio of
the cyclotron energy of the hole, $\hbar\omega_h\equiv\hbar e B/m_h$,
to the typical interaction energy scale,
$e^2/4\pi\epsilon\epsilon_0\ell$, is rather small, 0.15 (using
$\epsilon=12.53$ for GaAs), and one can expect Landau level mixing for
the hole to be quantitatively important.  Indeed, we will show that it
is the finite mass of the hole that provides the more important
mechanism by which the ``hidden symmetry'' is broken in the
photoluminescence spectrum.

The neglect of Landau level mixing for the electrons is the principal
assumption of our work and leads to the key simplifications.  It
allows explicit knowledge of the groundstate of the system at $\nu=1$:
a filled Landau level of spin-$\uparrow$ electrons\cite{sondhi}.
Moreover, we shall always consider interactions which preserve the
electron spin.  Therefore, prior to recombination of the valence-band
hole, the system may be characterized by the number of
spin-$\downarrow$ electrons and the number of {\it missing}
spin-$\uparrow$ electrons in the otherwise filled lowest Landau level
(``spin-holes'').  Through the use of this particle-hole
transformation, the initial states may described by the interaction of
the hole with (spin-$\downarrow$) electrons and spin-holes, both
restricted to states in the lowest Landau level.  For filling
fractions close to $\nu=1$, and for states which do not involve a
large degree of spin-depolarization, relatively few of these particles
are present, and the calculation of the energy eigenstates poses a
few-body problem.  The majority of our work will address the
properties of the system in which there are only two such particles;
our results in this case are based on analytical treatments.  We will
also present results of numerical studies for systems with larger
numbers of particles.  We work in the spherical
geometry\cite{haldanehierarchy,fanoortolani} and introduce the
separation $d$ between the electrons and hole in the same way as was
done in Ref.~\onlinecite{apalkovpikusefros}.  For all of the
calculations that we report, the system size is sufficiently large
that finite-size effects are well-controlled.

Although the neglect of Landau level mixing is not likely to be
quantitatively accurate at weak fields, when the typical interaction
energy can be larger than the electron cyclotron energy, we hope that
it does give the correct qualitative picture.  In
Sec.~\ref{subsec:mixing}, we will indicate the extent to which one can
trust the qualitative features of a model neglecting Landau level
mixing for the electrons, and in Sec.~\ref{subsec:positive} will
calculate some quantitative corrections arising from this mixing.

\section{Excitonic States}
\label{sec:exciton}

In this section we will consider the introduction of an electron-hole
pair to the groundstate at $\nu=1$.  We will show that, provided that
the distance $d$ of the hole from the electron gas is not too large,
the low-energy states may be described by a band of ``excitonic
states'', defined below.  We will further show that the recombination
of these excitonic states can account for the main feature of the
spectra presented in Fig.~\ref{fig:experiment}(b): a sharp
recombination line in the $\sigma_+$ polarization and an
asymmetrically broadened line in $\sigma_-$.  We will develop models
for the line shapes in the two polarizations; first for a system with
no disorder, and then taking account of the long-range disorder
arising from remote ionized donors.  We will compare the predictions
of these models with the experimental observations.

\subsection{Definition of the Excitonic States}

As we have explained above, the principal assumption throughout our
discussion is that the electrons are confined to the lowest Landau
level.  In this case, the groundstate of the system at $\nu=1$ prior
to photo-excitation is the state
\begin{equation} 
\label{eq:gstat} 
|0\rangle \equiv \prod_m e_{m\uparrow}^\dagger |\mbox{vac}\rangle ,
\end{equation}
in which the electrons fill all the spin-$\uparrow$ states in the
lowest Landau level and all spin-$\downarrow$ states are unoccupied.
In the above expression, $|\mbox{vac}\rangle$ is the vacuum state with
a filled valence band and empty conduction band, and
$e_{m\sigma}^\dagger$ is the operator which creates a spin-$\sigma$
electron in a single particle state in the lowest Landau level.  The
quantum number $m$ is any internal quantum number that runs over all
degenerate states in the lowest Landau level.  This state is clearly
the absolute groundstate if the bare electron Zeeman energy, $Z$, is
large compared to the typical interaction energy, set by
$e^2/4\pi\epsilon
\epsilon_0\ell$.
Due to the spontaneous ferromagnetism which appears for repulsive
electron-electron interactions, it is also the groundstate in the
limit $Z\rightarrow 0$\cite{sondhi}.

We now consider the introduction of an additional electron-hole pair
to the system.  The properties of the low-energy states depend on all
of the model parameters: $d/\ell$ and the ratios of the cyclotron
energy of the hole, $\hbar\omega_h$, and the electron Zeeman energy,
$Z$, to the typical interaction energy scale
$e^2/4\pi\epsilon\epsilon_0\ell$.  However, if the Zeeman energy is
large compared to the interaction energy, the low-energy states will
be maximally spin-polarized, and their form is clear: all of the
spin-$\uparrow$ electron states will be occupied, and there will be
one remaining spin-$\downarrow$ electron which will bind with the
valence-band hole to form an exciton (in a magnetic field, any
attractive interaction will lead to binding of a two-dimensional
electron-hole pair).  We refer to these states as ``excitonic
states''.  Since we assume that all interactions conserve the electron
spin and we ignore Landau level mixing for the electrons, the filled
Landau level of spin-$\uparrow$ electrons is inert, and the properties
of the excitonic states may be determined by considering only the
electron-hole pair.  In particular, the energy eigenstates of the
system follow from those of the exciton itself.  The state in which
the exciton is in a state $\Psi_P(\bbox{r}_e,\bbox{r}_h)$ with
momentum $\bbox{P}$ is
\begin{equation} 
\label{eq:excitonic} 
|\bbox{P}\rangle \equiv \int d^2\bbox{r}_e d^2\bbox{r}_h \;
\Psi_P(\bbox{r}_e,\bbox{r}_h) \psi_{e\downarrow}^\dagger(\bbox{r}_e)
\psi_{h}^\dagger (\bbox{r}_h)|0\rangle .
\end{equation}
We have suppressed the subband-label of the hole, but it is to be
understood that there are two excitonic bands corresponding to the two
hole states. In the absence of an external potential, the momentum
$\bbox{P}$ is conserved and the above states are energy eigenstates.
In appendix~\ref{app:exciton} the wavefunctions and dispersion
relation of a two-dimensional exciton are derived at small momentum as
a function of $m_h$ and $d$, within the approximation of no Landau
level mixing for the electron.

\subsection{Stability to Spin Reversal}
\label{subsec:stability}

The above excitonic states are the low-energy eigenstates of the
system when the Zeeman energy is large compared to the interaction
energy scale.  However, for typical GaAs samples at $\nu=1$, the
Zeeman energy of electrons is much smaller than the interaction energy
scale (at 4T the Zeeman energy is 0.09meV, whereas the typical
interaction energy is $e^2/4\pi\epsilon\epsilon_0\ell=8.9\mbox{meV}$).
It is therefore important, for practical purposes, to study whether
these excitonic states remain the lowest-energy states when the Zeeman
energy is small.  It is possible that there exist lower energy states
involving some degree of spin depolarization.  Such depolarization is
known to be important for the charged excitations of this system, for
which theoretical\cite{sondhi,fertigskyrmion,macdonaldskyrmion} and
some experimental\cite{barrett,schmeller,aifer} studies show that the
lowest energy charged excitations are ``charged spin-textures'' which
diverge in size to become ``skyrmions'' in the limit of vanishing
Zeeman energy.

The spin-polarization of the groundstate depends strongly on the
system parameters ($d$, $m_h$ and $Z$).  For sufficiently large $d$,
the system will become depolarized when $Z$ is small: in this case,
the interactions between the electron gas and the hole may be
neglected and the groundstate spin polarization will be that of the
extra electron, which is determined by the lowest-energy charged
spin-texture.  However, for small $d$ the hole is tightly bound to the
additional electron, and it is possible that the resulting neutral
exciton does not significantly disturb the spin-polarization of the
remaining electrons.

We have studied the stability of the exciton state to spin-reversal by
calculating the groundstate of the system in the presence of a {\it
single} spin-reversal.  Our calculations were performed in the
spherical geometry, with system sizes of up to 51 single particle
basis states in the lowest Landau level (corresponding to a sphere of
diameter $10\ell$).  We have studied only the limiting cases $m_h=0,
\infty$ and assume that the results for finite hole mass lie between
these two limits.  We find that, for $m_h=0$ and $m_h=\infty$, the
groundstate is the global spin-rotation of the excitonic groundstate
provided $d<1.4\ell$ and $d<1.3\ell$, respectively (we have identified
these values to an accuracy of better than $\pm0.1\ell$).  Therefore,
for $d$ smaller than these values, there is no energetic advantage to
be gained from introducing a single spin-flip to the excitonic state:
the zero-momentum excitonic state remains the groundstate.  In order
to fully test the stability of the excitonic state to spin-reversal,
one should study the groundstate as a function of all possible
spin-polarizations.  However, it seems likely that if the energy is
not reduced by the introduction of a single spin-reversal, it will not
be reduced by a larger depolarization.  We therefore anticipate that
for $d$ smaller than $1.3\ell$, the $\bbox{P}=0$ excitonic state is
the absolute groundstate of our model whatever value the hole mass may
take, and even as $Z\rightarrow 0$.  For larger values of $d$, the
spin-polarization of the groundstate will change as $Z$ is decreased;
one may view the resulting depolarized states as excitons formed from
the binding of a valence-band hole with a charged spin-texture.

For the parameter-values that we use to compare with the experiments
reported in Ref.~\onlinecite{plentz} the spacing $d=60\mbox{\AA}$ is
much less than the magnetic length $\ell = 130\mbox{\AA}$.  Our
calculations therefore suggest that in this sample the hole is
sufficiently close to the electron gas that the $\bbox{P}=0$ excitonic
state provides a good description of the groundstate of the system
exactly at $\nu=1$ prior to recombination.  We therefore expect the
excitonic states to provide an important contribution to the
photoluminescence spectrum of this sample.

At finite temperatures, some of the low-energy excited states of the
system will also be populated.  These will consist both of
finite-momentum exciton states, and of long-wavelength spin-wave
excitations of the system.  For small electron-hole separation, $d$,
and for sufficiently small excitation energies that the wavelengths of
these excitations are large compared to the magnetic length, the
coupling between the exciton and spin-waves will be small, and the two
may be treated independently.  In the remainder of this section we
will discuss the form of recombination expected from the excitonic
states in the absence of spin-waves.  A thermal population of
spin-waves may be viewed as a fluctuation in the overall polarization
of the system, and will lead to a mixing between the two circular
polarizations of the spectra.  Provided the temperature is small
compared to the particle-hole gap at $\nu=1$, only a small number of
spin-waves will be thermally populated, and this mixing will be small.

\subsection{Radiative Recombination}
\label{subsec:recombination}

\begin{figure}
\inseps{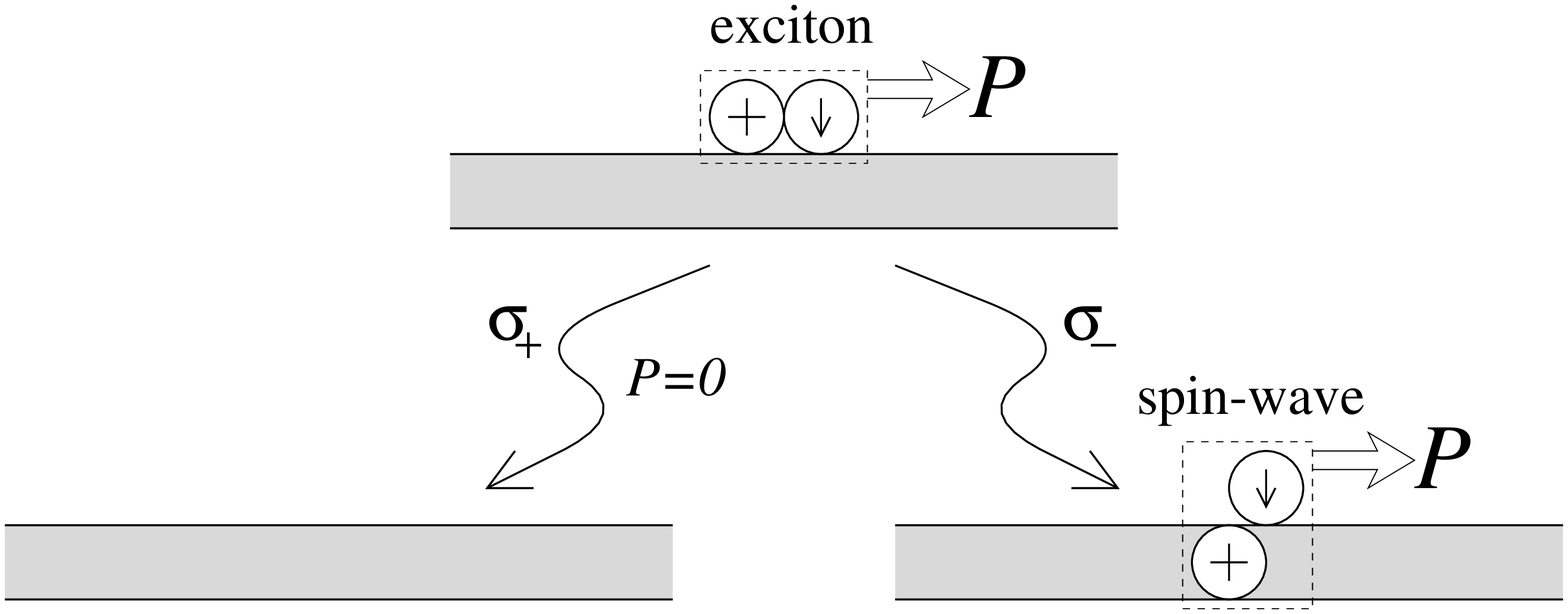}{0.3}
\vskip0.2cm
\caption{ Schematic diagram of the radiative decay processes
of the free excitonic states.  Each shaded region represents a filled
Landau level of spin-$\uparrow$ electrons, and the circles represent a
spin-$\downarrow$ electron (marked by $\downarrow$), a valence-band
hole (lying above the shaded region and marked by $+$), and a
``spin-hole'' (lying in the shaded region and marked by $+$).}
\label{fig:xtransitionsa}
\end{figure}

It is clear from the form of the excitonic states described above that
their radiative recombination in the two circular polarizations will
lead to quite different final states.  In the $\sigma_+$ polarization,
the hole (which must be in the $\Downarrow$ state) must recombine with
the single spin-$\downarrow$ electron.  In this case, there is only
one available final state: the $\nu=1$ groundstate (\ref{eq:gstat}).
In the $\sigma_-$ polarization, the hole can recombine with any one of
the spin-$\uparrow$ electrons and there are many possible final
states.  These are the states in which the spin of a single electron
has been reversed in the $\nu=1$ groundstate, and are described by the
set of single spin-wave excitations\cite{bychkov,kallinhalperin1}. The
recombination processes in the two polarizations are illustrated
schematically in Fig.~\ref{fig:xtransitionsa}.

The transition rates of these processes are determined by the matrix
elements of the operators, $L_\pm$.  Using the form of the exciton
wavefunctions derived in appendix~\ref{app:exciton}, these matrix
elements can be calculated explicitly.

We find that the matrix element describing $\sigma_+$ recombination
between the excitonic state $|\bbox{P}\rangle$ and the groundstate
$|0\rangle$ is
\begin{equation}
\label{eq:xmatrixelement}
\langle 0|L_+|\bbox{P}\rangle = \sqrt{\frac{\Omega}{2\pi \ell^2}}
\;\delta_{P,0},
\end{equation}
where $\Omega$ is the area of the sample.  Thus, on emission of a
long-wavelength photon, momentum conservation limits the recombination
to the zero-momentum excitonic state.

\begin{figure}
\inseps{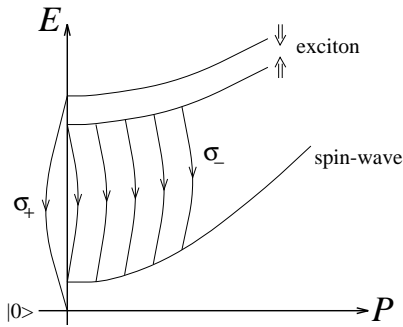}{0.3}
\vskip0.1cm
\caption{Schematic diagram of the allowed radiative transitions
of the free excitonic states as a function of momentum. The two
exciton bands arise from the two hole subband states, and give rise to
the two circular polarizations of emitted radiation.  In the
$\sigma_+$ polarization only the zero-momentum state recombines and
leaves the groundstate $|0\rangle$ as the final state; in the
$\sigma_-$ polarization (vertical) transitions occur from excitonic
states of all momenta into the band of spin-wave states.}
\label{fig:xtransitionsb}
\end{figure}

As we now show, for the $\sigma_-$ polarization {\it all} of the
excitonic states can recombine, with the momentum of the exciton being
conserved by the momentum of the spin-wave in the final state.  Making
a particle-hole transformation on the filled Landau level of
spin-$\uparrow$ electrons, the matrix element of $L_-$ between an
initial excitonic state, $|\bbox{P}\rangle$, and a final state
$|\bbox{P}^\prime\rangle_{SW}$ in which there is a single spin-wave
with momentum $\bbox{P}^\prime$, may be written
\widetext
\begin{equation}
\label{eq:xswmatrixelement}
_{SW}\langle \bbox{P}^\prime |L_-|\bbox{P}\rangle = \int
\Psi_{SW,\bbox{P}}^*(\bbox{r}_e,\bbox{r}_h)\Psi_{X,\bbox{P}}
(\bbox{r}_e,\bbox{r}_h)
\;d^2\bbox{r}_e d^2\bbox{r}_h,
\end{equation}
where $\Psi_{X,P}(\bbox{r}_e,\bbox{r}_h)$ and
$\Psi_{SW,P}(\bbox{r}_e,\bbox{r}_h)$ are the wavefunctions of the
exciton and spin-wave of momentum $\bbox{P}$, with $\bbox{r}_h$
representing the position of the spin-hole in the second case.  In the
symmetric gauge, $\bbox{A}(\bbox{r})=
B\hat{\bbox{z}}\times\bbox{r}/2$, the exciton and spin-wave
wavefunctions may be written\cite{gorkov}
\begin{equation}
\Psi_P(\bbox{r}_e,\bbox{r}_h)= \frac{1}{\sqrt{\Omega}}
e^{i\bbox{P}.(\bbox{r}_e+\bbox{r}_h)/2\hbar}e^{i\bbox{r}_e\times
\bbox{r}_h.
\hat{\bbox{z}}/2\ell^2}
\Phi_{\bbox{P}}(\bbox{r}_e-\bbox{r}_h).
\end{equation}
Integrating (\ref{eq:xswmatrixelement}) over the co-ordinate,
$(\bbox{r}_e+\bbox{r}_h)/2$, we obtain
\begin{equation}
\label{eq:xswmatrixelementint}
_{SW}\langle \bbox{P}^\prime |L_-|\bbox{P}\rangle =
\delta_{P,P^\prime}
\int \Phi_{SW, P}^*(\bbox{r})\Phi_{X, P}(\bbox{r})\; d^2\bbox{r},
\end{equation}
\bottom{-2.7cm}
\narrowtext
\noindent
which demonstrates that the transition from the excitonic to the
spin-wave state occurs with momentum conservation, and at a rate
depending on the overlap of their respective internal wavefunctions.
The spin-wave wavefunctions $\Phi_{SW, P}$ are
well-known\cite{lernloz,kallinhalperin1}. They are fully specified by
the condition that both electron and hole (i.e. missing
spin-$\uparrow$ electron) are in the lowest Landau level, and are
independent of the force-law between the electrons.  Since we allow
Landau level mixing for the valence-band hole, the exciton
wavefunctions do depend on the strength of the interaction relative to
the cyclotron energy of the hole.  However, as shown in
appendix~\ref{app:exciton}, for small momenta the wavefunction of the
exciton is identical to that of the spin-wave, and we obtain
\begin{equation}
\label{eq:xswmatrixelementfinal}
_{SW}\langle \bbox{P}^\prime |L_-|\bbox{P}\rangle =
\delta_{P,P^\prime} + {\cal O}( P^2\ell^2/\hbar^2 ),
\end{equation}
independent of the parameters $m_h$ and $d$.  (The corrections at
finite momentum vanish for all $d$, when $m_h\rightarrow 0$, in which
limit Landau level mixing of the hole is negligible and the exciton
wavefunction becomes identical to that of the spin-wave.)  The allowed
recombination processes for both $\sigma_+$ and $\sigma_-$
polarizations are illustrated in Fig.~\ref{fig:xtransitionsb} as a
function of momentum.

\subsection{Line Shapes: No Disorder}

The fact that only the $\bbox{P}=0$ excitonic state can contribute to
photoluminescence in the $\sigma_+$ polarization, whereas all of the
excitonic states can contribute to emission in the $\sigma_-$
polarization leads to quite different line shapes for the two
polarizations.  It is immediately apparent that, in the absence of
disorder, the $\sigma_+$ emission must be a sharp line, since there is
only one possible transition. In the $\sigma_-$ polarization many
transitions can occur and one can expect to observe a broad line.

To understand the line shape of the $\sigma_-$ polarization, consider
first the case $m_h=0$, $d=0$, in which there is no Landau level
coupling for the valence-band hole, and it lies in the same plane as
the electrons.  In this case, the dispersion relations of the exciton
and the spin-wave are {\it identical}, so all of the allowed
$\sigma_-$ transitions have the same energy.  This situation provides
an illustration of how the ``hidden symmetry'' that applies in this
case ($m_h=0$, $d=0$) leads to a trivial spectrum in which all
transitions within each polarization occur at the same energy.

In order to observe structure in the $\sigma_-$ recombination line, it
is essential that the dispersion relations of the exciton and
spin-wave differ.  Differences arise {\it both} from a non-zero value
of $d$, such that the electron-hole interaction differs from the
electron-electron interaction, and from Landau level mixing of the
hole states. (Differences will also arise from Landau level mixing of
the electrons, but these effects lie outside the scope of the theory
presented here.)  We will compare the dispersion relations of the
excitonic and spin-wave states by discussing the effective masses of
these two excitations, which describe the properties at small momenta.

In appendix~\ref{app:exciton} we show that, within a model that
neglects Landau level mixing for the electron, the effective mass of
the exciton, $M_X$, may be calculated exactly
\widetext
\begin{equation}
\label{eq:excitonmass}
M_X=m_h + \frac{4\pi\epsilon\epsilon_0\hbar^2}{e^2\ell}
\left[\sqrt{\frac{\pi}{8}}\exp{(d^2/2\ell^2)}(1+d^2/\ell^2)\mbox{erfc}
(d/\sqrt{2}\ell) -d/2\ell\right]^{-1}.
\end{equation}
\bottom{-2.7cm}
\narrowtext
\noindent
The effective-mass approximation to the exciton dispersion relation is
good for $|\bbox{P}|\ll (1+\lambda)\hbar/\ell$, where $\lambda\ge 0 $
is a parameter defined in Eq.(\ref{eq:lambda}) that describes the
extent of Landau level mixing for the hole.  The spin-wave dispersion
relation is parabolic for $|\bbox{P}|\ll \hbar/\ell$ and may be
described by an effective mass\cite{bychkov,kallinhalperin1}
\begin{equation}
M_{SW}= \sqrt{\frac{8}{\pi}}\frac{4\pi\epsilon\epsilon_0\hbar^2}{e^2\ell},
\end{equation}
which may be obtained from (\ref{eq:excitonmass}) by setting $m_h$ and
$d$ to zero.  From equation (\ref{eq:excitonmass}) we find that the
Landau level mixing and the spatial separation of the hole from the
electron gas both {\it increase} the effective mass of the exciton
relative to that of the spin-wave.  For the parameter-values
appropriate to the sample used in Fig.~\ref{fig:experiment}(b) we find
an exciton effective mass of $M_X=0.50m_0$, which is much larger than
that of the spin-wave, $M_{SW}=0.081m_0$.  Most of this increase
arises from the finite mass of the hole; ignoring Landau level mixing
for the hole ($m_h=0$), one would estimate $M_X=0.16m_0$.  For this
sample, therefore, we find that Landau level mixing of the hole
provides the more significant mechanism by which the ``hidden
symmetry'' in photoluminescence is broken.  Even in a stronger
magnetic field of $8\mbox{T}$, when Landau level mixing effects are
less important, one finds that a very large value of the spacing,
$d\gtrsim\ell$, is required before the contribution to the mass
difference between the exciton and spin-wave arising from the non-zero
$d$ outweighs that due to the Landau level mixing of the hole.  Thus,
it is typically the case that Landau level mixing for the hole
provides a more important contribution to the loss of the ``hidden
symmetry'' than the spacing $d$ in the spectrum arising from the
excitonic states.

We claim that the observations reported in Ref.\onlinecite{plentz} and
reproduced in Fig.~\ref{fig:experiment}(b) demonstrate the
recombination of excitonic initial states.  This claim is motivated
both by the discussion of Sec.~\ref{subsec:stability} in which
it was shown that the excitonic states are the low-energy initial
states of our model using the parameter-values appropriate for the
experiments, and also by the fact that the principal qualitative
features of the observed spectra are consistent with the those
expected from the recombination of excitonic states.  Namely, in the
$\sigma_+$ polarization there is a sharp recombination line, due to
the recombination of the $\bbox{P}=0$ excitonic state; while in
$\sigma_-$ one expects a broadening of the emission line on the {\it
low} energy side, due to the recombination of excitonic states with
non-zero momentum and the subsequent shake-up of spin-waves.

\begin{figure}
\inseps{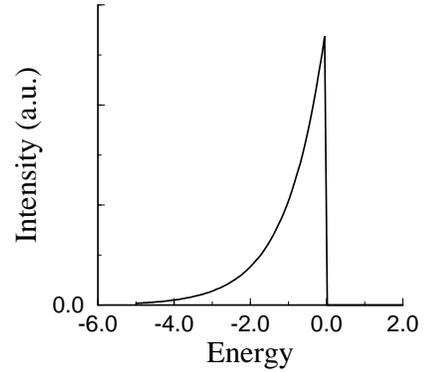}{0.5}
\vskip0.2cm
\caption{Line shape for the $\sigma_-$ polarized
luminescence from free excitonic states in thermal equilibrium at a
temperature $T$, assuming parabolic bands for both exciton and
spin-wave.  The energy is measured in units of $k_B T(M_X/M_{SW} -1)$,
relative to the recombination energy of the zero-momentum exciton in
this polarization.}
\label{fig:xtransitionsc}
\end{figure}

To make closer comparison between our theory and the experiments, we
proceed by calculating line shapes for the excitonic recombination. To
do so, it is essential to know the relative probabilities, $n_P$, for
occupation of the various excitonic states.  Assuming that these are
populated according to a Boltzmann distribution at a temperature $T$
and treating the exciton and spin-wave dispersions as parabolic, it is
straightforward to show that the line shape is
\begin{eqnarray}
I_-(E) & \equiv & \sum_{\bbox{P}}
n_{\bbox{P}}\; \delta\left[E-(E^X_P-E^{SW}_P)\right] \\
\label{eq:thermalline}
 & \propto &   
\exp\left[{\frac{E}{k_BT(M_X/M_{SW}-1)}}\right] \Theta\left( -E\right) ,
\end{eqnarray}
where the recombination energy is measured relative to that of the
$\bbox{P}=0$ excitonic state in this polarization, and we have assumed
$M_X>M_{SW}$.  The resulting line shape is illustrated in
Fig.~\ref{fig:xtransitionsc}.  The recombination in the $\sigma_+$
polarization is insensitive to the differences between the spin-wave
and exciton dispersion relations, and remains a single sharp line.

The line shape in Fig.~\ref{fig:xtransitionsc} is similar to the form
of the main recombination line observed experimentally in the
$\sigma_-$ polarization [Fig.~\ref{fig:experiment}(b)].  Furthermore,
the line width predicted by the above theory is comparable to, though
slightly less than, the width of the experimental spectrum: within our
approximations, the ratio of the exciton to spin-wave effective mass
is $M_X/M_{SW}=6.2$, so, at the experimental temperature of 0.5K, we
have $k_BT(M_X/M_{SW}-1)=0.2\mbox{meV}$ (at this temperature, the
thermal wavelength of the exciton is $2600\mbox{\AA}$, which is large
compared to the magnetic length, $\ell=130\mbox{\AA}$, so the
effective mass approximation is accurate for both exciton and
spin-wave dispersions).  However, there is a qualitative discrepancy
between the above expression (\ref{eq:thermalline}) and the
experimentally-observed line shape.  Namely, the width of the observed
line does {\it not} change with temperature for temperatures below the
$\nu=1$ single particle gap (above this temperature, peculiarities
associated with the $\nu=1$ filling fraction do not appear in the
experimental spectra).  This discrepancy indicates a failure of the
above model for the line shape of the excitonic states.  In the
following subsection we show that the assumption of {\it free}
excitonic states is not likely to be accurate in these systems:
disorder can be expected to lead to strong scattering of the excitonic
states.  Including the effects of disorder, we show that the excitonic
recombination spectrum develops a temperature-independent line shape
that is consistent with the experimental observations.

\subsection{Effects of Long-Range Disorder}
\label{subsec:disorder}

There are various sources of disorder which can scatter the excitons
discussed above.  These include interface roughness and impurities
that scatter excitons in undoped quantum wells.  However, an
additional source of disorder appears in modulation-doped quantum
wells: the long-range potential arising from the donor impurities that
lie some distance, set by the spacer-layer thickness, from the quantum
well.  (The exciton will also interact with the quasiparticles that
appear when the filling fraction is not exactly $\nu=1$.  See
Sec.~\ref{sec:charges}.)  The long-range potential fluctuations are
believed to be the dominant source of exciton scattering in the sample
of Ref.~\onlinecite{plentz}, the effects of interface roughness being
small due to the large width and asymmetry of the quantum
well\cite{flavio}. In this section we will discuss the recombination
of the excitonic states in the presence of the long-range potential
disorder, by studying the energy eigenstates of both the exciton and
the spin-wave (the final state of the $\sigma_-$ recombination
process).  By including this single source of disorder, we obtain a
lower limit on the extent of the disorder-broadening of the spectral
lines.

We begin by reviewing the form of the disorder arising from the
ionized donors, as has been discussed in
Refs.\onlinecite{kukushkindos} and~\onlinecite{pikusefros}.  Imagine
that these donors lie in a plane a distance $s$ (the spacer-layer
thickness) from the two-dimensional electron gas, with an average
density $n$ (the same as the number density of the two-dimensional
electron gas).  If the donors are randomly distributed in the plane,
then the density fluctuations $n_q$ are correlated according to
\begin{equation}
\label{eq:donorcorrelation}
\overline{n_q n_{-q^\prime}} = n \Omega\delta_{q,q^\prime}
\end{equation}
where $\Omega$ is the sample area and the bar indicates the average
over all realizations of the disorder.  Due to their mutual
electrostatic repulsion, one expects there to be significant
correlations between the positions of the ionized donors, and a
subsequent reduction in the amplitude of the density fluctuations.  We
treat these correlations within the ``non-equilibrium model'' of
Ref.\onlinecite{pikusefros}, in which the donor distribution is
assumed to be a snapshot of the distribution at a temperature $T_0$.
This model is based on the idea that, as the sample is cooled, the
charges on the ionized donors readjust within the impurity band until
a temperature $T_0$ is reached at which such charge-mobility becomes
small ($T_0\simeq 100\mbox{K}$ typically).  It is found that, on
lengthscales larger than the spacer thickness, the correlation
function for the donor fluctuations takes the same form as
Eq.(\ref{eq:donorcorrelation}), but with the density replaced by an
effective density $n^*=\epsilon\epsilon_0 k_B T_0/(e^2 s)$.

The fluctuations of the donor density lead to fluctuations in the
potentials experienced by the electrons and holes.  The resulting
potential energy fluctuations for the electrons are
\begin{equation}
\label{eq:ve}
V^e_q = \frac{-e^2}{2\epsilon\epsilon_0|\bbox{q}|} n_q
e^{-|\bbox{q}|s},
\end{equation}
and are therefore suppressed on scales larger than the spacer layer
thickness, $s$.  Due to the asymmetry of the quantum well, the
centre-of-charge of the hole is located a distance $d$ further from
the ionized donors than that of the electrons, so the magnitude of the
potential experienced by the hole is slightly smaller
\begin{equation}
\label{eq:vh}
V^h_q = -V^e_q
e^{-|\bbox{q}|d}\stackrel{d\ll s}{\simeq} -V^e_q(1-|\bbox{q}|d),
\end{equation}
but is also smooth on a lengthscale $s$.

In appendix~\ref{app:exciton} we have derived an effective Hamiltonian
for the motion of the exciton in smooth external potentials
$V^e(\bbox{r})$ and $V^h(\bbox{r})$. We make a Born-Oppenheimer
approximation for the exciton motion, treating the internal motion
(with a characteristic timescale set by the exciton energy level
separation) as fast compared to the scattering rate of the exciton by
the potential.  Expanding the potentials to lowest order in $\ell /s$,
where $s$ is the lengthscale of the external potentials, the effective
Hamiltonian for the centre-of-mass position $\bbox{R}$ and momentum
$\bbox{P}$ of the exciton is found to be
\widetext
\top{-2.8cm}
\begin{equation}
H^{eff}_X =  \frac{\bbox{P}^2}{2M_X}
+ V^e(\bbox{R}) + V^h(\bbox{R})
+ \frac{1}{1+\lambda}\frac{\ell^2 \bbox{P}}{\hbar}.
(\eta_h\nabla V^e - \eta_e\nabla V^h)\times\hat{\bbox{z}}
  - \frac{\lambda}{2(1+\lambda)}\frac{\ell^2}{2}
 \frac{\left| \eta_h\nabla V^e - \eta_e
\nabla V^h\right|^2}{V^{eh}_1-V^{eh}_0} ,
\label{eq:excitonhamiltonian}
\end{equation}
\bottom{-2.7cm}
\narrowtext
\noindent
where all gradient operators are to be understood to act in the plane
of the quantum well.  The first term represents the kinetic energy of
the exciton, with an effective mass, $M_X$, given by
Eq.(\ref{eq:excitonmass}), while the second and third terms describe
the potential energy of an exciton with centre-of-mass position
$\bbox{R}$.  The fourth term, in which $\eta_{e,h}\equiv
m_{e,h}/(m_e+m_h)$ and $\lambda$ is the parameter defined in
Eq.(\ref{eq:lambda}), represents the coupling of the in-plane dipole
moment of the exciton with the electric fields acting on the electron
and hole parallel to this plane. This term should be symmetrized in
momentum and position co-ordinates to render the Hamiltonian
Hermitian. The final term, in which $V^{eh}_m$ are expectation values
of the electron-hole interaction defined in
appendix~\ref{app:exciton}, is the Stark shift of the exciton in the
parallel components of the electric field due to the mixing with the
higher exciton bands.

The above Hamiltonian also describes the motion of the spin-waves of
the $\nu=1$ state in the long-range potential.  In this case, the hole
represents the spin-hole and one must therefore restrict it to states
in the lowest Landau level ($\lambda\rightarrow 0$) and use
$V^h(\bbox{r})=-V^e(\bbox{r})$.  The effective spin-wave Hamiltonian
therefore reduces to
\begin{eqnarray}
H^{eff}_{SW} & = & \frac{\bbox{P}^2}{2M_{SW}}
+ \frac{\ell^2 \bbox{P}}{\hbar}.\left(\nabla V^e \times\hat{\bbox{z}}\right)\\
 & = & \frac{\left(\bbox{P}-q^*\bbox{A}^*\right)^2}{2M_{SW}} +
 {\cal O}(|\nabla V^e|^2) .
\end{eqnarray}
\noindent
Thus, the disorder couples only through the in-plane dipole moment of
the spin-wave.  In the last line, we have rewritten the Hamiltonian in
the more familiar form of the free motion of a particle with
(fictitious) charge $q^*$ in a random (fictitious) vector potential,
$\bbox{A}^*\equiv -M_{SW}\ell^2 /(q^*\hbar) \nabla
V^e\times\hat{\bbox{z}}$, neglecting a second-order term in the electric
field.  The effective magnetic field experienced by the particle is a
random function of position with zero average $B^* =\nabla\times
\bbox{A}^* =\nabla^2 V^e M_{SW}\ell^2 /(q^*\hbar)$.  The effect of this
magnetic field on the motion of the particle may be judged by
considering the typical (fictitious) magnetic length.  Calculating the
root mean square value of $\nabla^2 V^e$ for the donor distribution
(\ref{eq:donorcorrelation}) with the reduced density $n^*$, we find a
typical magnetic length
$\ell^*=s\sqrt{(\hbar^2/M_{SW}\ell^2)(8\sqrt{\pi}\epsilon\epsilon_0/\sqrt{3
n^*}e^2)}$.  Using the parameter-values $T_0=100\mbox{K}$, $\ell =
130\mbox{\AA}$, $s=800\mbox{\AA}$, which are appropriate for the
sample of Ref.~\onlinecite{plentz} under the conditions for which
Fig.~\ref{fig:experiment}(b) was measured, we find the typical
effective magnetic length is approximately twice the disorder
lengthscale $s$.  Thus the radius of curvature of the spin-wave
trajectory is always large compared to the disorder lengthscale, and
the scattering by this random magnetic field is {\it weak}. We will
neglect the effects of this weak disorder on the spin-wave motion, and
treat the spin-wave as a free particle with a parabolic dispersion
relation described by $M_{SW}$.

The strength of scattering of the exciton due to the coupling of its
in-plane dipole moment to the in-plane electric field is of the same
order as the scattering of the spin-wave, and is therefore also small.
However, the scattering arising from the remaining terms is strong.
The main contribution is due to the potential $V^X(\bbox{r})\equiv
V^e(\bbox{r})+V^h(\bbox{r})$, which describes the coupling of the
perpendicular dipole moment of the exciton to the fluctuations in the
electric field.  Using the correlation function
(\ref{eq:donorcorrelation}) and the expressions (\ref{eq:ve}) and
(\ref{eq:vh}) for the electron and hole potential energies, we find
that the root mean-square fluctuation in the exciton energy is
$V^X_{rms}= 0.093\mbox{meV}$ for the parameter-values of the sample of
Ref.~\onlinecite{plentz} ($s=800\mbox{\AA}$, $d=60\mbox{\AA}$ and
assuming $T_0=100\mbox{K}$).  Since the fluctuation in the exciton
energy is large compared to the kinetic energy cost $\sim \hbar^2/M_X
s^2= 0.02\mbox{meV}$ to confine it to a region of size $s$, one
expects the disorder to lead to strongly-localized low-energy states.

In view of the above considerations, we arrive at a model for the
excitonic recombination in which the exciton states prior to
recombination must be determined from the potential $V^X(\bbox{r})$,
and the final states are either the groundstate (in the $\sigma_+$
polarization) or the {\it free} spin-wave ($\sigma_-$). To derive line
shapes for the resulting photoluminescence spectra, one must also know
the relative populations of the initial exciton states.

One possibility is to assume that thermal equilibrium is achieved.  In
this case, the width of the $\sigma_+$ line, in which the exciton
recombines to leave the $\nu=1$ groundstate, will vanish as the
temperature tends to zero and the exciton becomes restricted to its
low-energy ``tail
states''\cite{halperinlax1,halperinlax2,lifshitz1,lifshitz2}. However,
the line shape in the $\sigma_-$ polarization will be much broader
than in $\sigma_+$ as a result of the shake-up of high-momentum
spin-waves.  In fact, at very low temperatures, this line shape will
adopt a temperature-independent form, determined by the
momentum-distribution of the expected limiting form of the tail-state
wavefunctions\cite{halperinlax1,halperinlax2}. We do not pursue a
calculation of this line shape, since we do not believe that this
limiting behaviour is appropriate for the present experiments.
Rather, as is the case for exciton recombination in empty quantum
wells\cite{noneqexciton,yang}, we expect that the finite lifetime of
the valence-band hole prevents full thermal equilibration.  This is
consistent with the observed temperature-independent width of the
$\sigma_+$ line\cite{plentz}. Moreover, in view of the above estimates
for the strength of disorder which show that the low-energy exciton
states are likely to be strongly-localized in the potential minima,
one might expect a slow equilibration rate.

\begin{figure}
\inseps{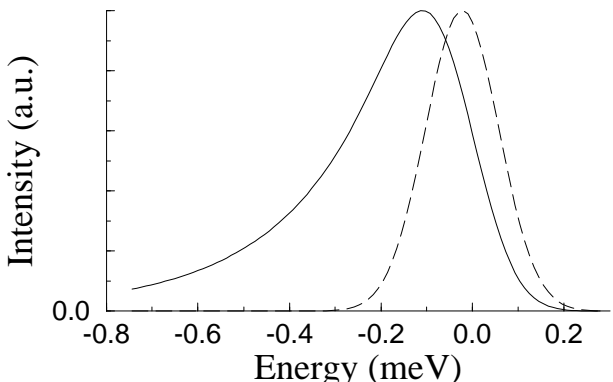}{0.6}
\vskip 0.1cm
\caption{Spectra of recombination of the
excitonic states in the presence of long-range diso<rder, for
$\sigma_-$ (solid line) and $\sigma_+$ (dashed line) polarizations.
Energies are measured relative to the recombination of the
zero-momentum exciton in a sample without disorder.  Parameter-values
are chosen for the sample of Fig.~\protect\ref{fig:experiment}(b):
$(\hbar^2/M_X s^2)/V^X_{rms} =0.25$, $M_X/M_{SW}=6.2$, $\hbar^2/M_X
s^2=0.024\mbox{meV}$.}
\label{fig:disorderlineshapes}
\end{figure}

If thermal equilibrium is not achieved, the line shapes will depend
both on the nature of the exciton states in the presence of disorder
and on the relaxation dynamics.  Since we do not have a good
understanding of the relaxation dynamics, we will treat the
non-equilibrium recombination within a very simplified model.  We
imagine that the disorder is sufficiently strong
($V^X_{rms}\gg\hbar^2/M_X s^2$) that the exciton can become strongly
confined in any local minimum of the potential, and that the tunneling
rate between states in different minima is small compared to the decay
rate of the valence-band hole.  Under these conditions, the low-energy
exciton states in all such minima accurately represent a set of energy
eigenstates, each of which will contribute to photoluminescence if
populated.  We represent the relaxation dynamics of the exciton by the
assumption that, prior to radiative recombination, the exciton is {\it
equally likely} to be found in any one the potential minima and that
only the groundstate in any given minimum is populated.  This
assumption is chosen to portray a rapid relaxation to the groundstate
in a given potential minimum and a slow equilibration between states
in different minima.  The same assumption was the key element of a
model proposed in Ref.~\onlinecite{yang} to account for exciton
recombination in an undoped quantum well.  Also in common with that
work, we assume that the potential is Gaussian-correlated; this is
accurate when the spacer layer thickness is large compared to the mean
impurity separation, such that many impurities contribute to the
potential at a given point of the two-dimensional electron gas.
Following a similar approach to that described in
Ref.~\onlinecite{yang}, we calculate the mutual probability
distribution of the potential $V^X$ and its curvatures in the two
principal directions at each point where $\nabla V^X=0$.  We use this
to calculate the spectra for the two polarizations by averaging over
the recombination of the exciton groundstate in all potential wells
(points of zero potential gradient for which both curvatures are
larger than zero), giving equal weight to each of these states.
Details of these calculations are presented in
appendix~\ref{app:disorder}.  The resulting line shapes are shown in
Fig.~\ref{fig:disorderlineshapes} for the parameter-values appropriate
for the conditions of Ref.~\onlinecite{plentz}.

In the $\sigma_+$ polarization, the radiative recombination of a given
exciton state contributes a sharp spectral line at an energy equal to
the value of the potential energy at the given potential minimum plus
the kinetic energy of the exciton.  For the parameter-values used in
Fig.~\ref{fig:disorderlineshapes}, the kinetic energy of the exciton
is small compared to the fluctuations in the potential, and the
overall linewidth is primarily determined by the width of the
distribution of the potentials at all minima, $\sim V^X_{rms}$.

In the $\sigma_-$ polarization, the recombination of each excitonic
state is broadened to low energy due to the shake-up of free
spin-waves with effective mass $M_{SW}$.  The extent of this
broadening depends on the momentum-distribution of the initial exciton
wavefunction.  The typical broadening may be estimated by considering
the groundstate wavefunction in a typical potential minimum, with
curvatures set by the root-mean-square curvature $\nabla^2 V^X_{rms}$.
The width of the emission line arising from this state is found to be
equal to $(M_X/M_{SW})\sqrt{\hbar^2 \nabla^2 V^X_{rms}/(2 M_X)}$.
This energy, which is $0.4\mbox{meV}$ for the parameter-values we use
to describe the experiments of Ref.~\onlinecite{plentz}, accounts for
the main low-energy broadening of the $\sigma_-$ spectrum shown in
Fig.~\ref{fig:disorderlineshapes}.

The theoretical line shapes shown in Fig.~\ref{fig:disorderlineshapes}
are very similar to the experimental line shapes
[Fig.~\ref{fig:experiment}(b)], both qualitatively and quantitatively.
It is important to emphasize, however, that the quantitative
predictions of our model are rather unreliable: the values of the
exciton and spin-wave masses we use do not take account of Landau
level mixing for the electrons, and are based on a simplified model
for the subband structure and the valence band dispersion.
Furthermore, our model for the exciton relaxation and recombination in
the disordered potential is only a crude description of a rather
complicated process.  An accurate calculation of the line shapes
requires a much better understanding of the relaxation dynamics of the
system than we have at present.  Consequently, the predictions of this
model are best viewed as illustrations of the qualitative features one
expects of the spectrum when the relaxation dynamics prevent full
thermal equilibration.  Specifically, the line shapes in both
polarizations become temperature-independent, and the $\sigma_-$
polarization is significantly broadened to lower energy as a result of
the release of high-momentum spin-waves upon recombination.  Finally,
we note that our calculations have dealt only with long-range
disorder.  Short-range disorder, arising for example from interface
roughness, may lead to high-momentum components in the exciton
wavefunction and have important quantitative effects on the width of
the $\sigma_-$ line. The qualitative features of the spectra will
remain the same.

\subsection{Effects of Landau Level Mixing}
\label{subsec:mixing}

To conclude this section on the recombination of excitonic states, we
will briefly discuss the validity of the approximation in which Landau
level mixing for the electrons is neglected.  This approximation is
correct in the limit in which the electron cyclotron energy
$\hbar\omega_e$ is large compared to the interaction energy scale
$e^2/4\pi\epsilon\epsilon_0\ell$.  For typical samples, these energies
are of a similar size, so one must always expect Landau level mixing
to have significant quantitative effects.  One may, however, hope that
the qualitative behaviour is correctly captured by such a theory.

The great deal of theoretical and experimental work on the integer and
fractional quantum Hall regimes has shown that this is the case for
many properties of these two-dimensional electron
systems\cite{prangeandgirvin}.  In particular, the prediction of a
spontaneously spin-polarized $\nu=1$ state with a particle-hole gap
determined largely by interactions\cite{sondhi} appears to be realized
in experiment\cite{barrett,schmeller,aifer}.

Similarly, we expect that a small amount of Landau level mixing will
not affect the qualitative properties of the excitonic states.
Corrections will arise from the coupling of these states with the
plasmons of the $\nu=1$ state.  For weak Landau level mixing, this
coupling leads to a weak-coupling polaronic problem in which a
particle (the exciton) couples to the density fluctuations of its
environment (the plasmons).  The parameter-values of the resulting
polaronic problem are such that one expects the energy eigenstates to
be closely related to the states in the absence of Landau level
mixing, only dressed with a cloud of virtual plasmons.

Thus, a small amount Landau level mixing will not lead to any
qualitative changes in the nature of the initial or final states of
the recombination process.  Quantitative changes will appear as
increases in the effective masses of the spin-wave and the exciton.
We know of no calculations of the effects of Landau level mixing on
these dispersion relations.  However, the form of the perturbation
expansion in the ratio of interactions to the electron cyclotron
energy shows that the lowest-order corrections to both effective
masses are proportional to $m_e$.  There will also be changes in the
matrix elements (\ref{eq:xmatrixelement}) and
(\ref{eq:xswmatrixelementfinal}).  However, these overlaps will still
vary on the characteristic momentum scale $\hbar/\ell$, so the
corrections will only lead to a uniform change of the recombination
rate for all excitonic states with small momenta.  Provided the mass
of the exciton remains larger than the mass of the spin-wave, all of
the above qualitative discussion will still apply.

It is possible that under the experimental conditions the extent of
Landau level mixing for the electrons is so large that the nature of
the low-lying initial states is qualitatively different: for example,
the spin polarization of the groundstate may be different.  Without a
full calculation of the many-body problem including Landau level
mixing, we cannot rule out such a possibility.  However, we view the
success of our model in explaining the main features of the
experimental observations as evidence that we have correctly
identified the states contributing to photoluminescence in these
experiments.  Experimentally, one may test whether Landau-level mixing
has any qualitative effects by studying the evolution of the spectrum
as a function of the sample density.  One would hope that the
qualitative features remain the same in higher density samples (but
with a similar value of $d/\ell$), for which $\nu=1$ appears at larger
magnetic field and Landau level mixing is less important.

\section{Charged initial States}
\label{sec:charges}

We have shown that exactly at $\nu=1$, the low-energy initial states
of our model are well described by excitonic states when $d$ is
sufficiently small.  However, this situation represents only a
singular value of the filling fraction.  For any typical filling
fraction close to $\nu=1$, the sample will contain a small number of
quasiparticles.  As the magnetic field increases and the average
filling fraction sweeps through unity, the quasiparticles will change
from being negative ($\nu\gtrsim 1$) to positive ($\nu\lesssim 1$).
These charges may become localized by disorder, in which case they
will not contribute to the transport properties and a quantized Hall
effect will be observed.  However, even localized charges may affect
the photoluminescence spectrum.  To discuss the consequences, we will
consider cases in which the filling fraction is sufficiently close to
one that the quasiparticles are very dilute (compared to the density
of electrons) and can be considered to be independent: we will
therefore study a single quasiparticle in an otherwise uniform $\nu=1$
state. We will consider a sample without disorder. Long-range disorder
will cause a broadening of all the spectra we describe below by an
amount similar to that of the exciton recombination line in the
$\sigma_+$ polarization ($\sim V^X_{rms}$), which was discussed in
Sec.~\ref{subsec:disorder}.

When an electron-hole pair is added to a system containing a single
additional quasiparticle, the resulting energy spectrum will contain a
band of states describing the motion of an exciton far from the
quasiparticle.  The recombination of these states is well described by
the discussion of Sec.~\ref{sec:exciton}, with the quasiparticle
providing an additional source of scattering.  However, it may be that
the groundstate does not form part of this band, but is some ``bound
state'' in which the additional electron-hole pair is localized in the
vicinity of the quasiparticle to form a small charged complex. If this
is the case, one can expect to find a separate feature in the
photoluminescence spectrum arising from this new initial state. The
simplest forms of these initial states (those with maximal
spin-polarization), were discussed in Ref.~\onlinecite{muzykantskii}.
We will also limit our discussion to the maximally spin-polarized
charged complexes, but extend the work of
Ref.~\onlinecite{muzykantskii} by showing that Landau level mixing for
the {\it electrons} can have important effects on the stabilities of
these complexes relative to the excitonic states, and that Landau
level mixing for the {\it hole} can significantly affect their
recombination spectra.

\subsection{Additional positive charge}
\label{subsec:positive}

We begin by considering a sample in which, prior to photo-excitation,
there is a single positively-charged excitation.  For large Zeeman
energy, the groundstate prior to photo-excitation is maximally
spin-polarized, and the positive charge appears as a vacant
spin-$\uparrow$ electron state (a ``spin-hole'').  In the absence of
disorder, the groundstate is any one of the degenerate states
\begin{equation}
\label{eq:spinhole}
e_{m\uparrow} |0\rangle ,
\end{equation}
classified by the quantum number $m$ describing the position of the
spin-hole in the plane.  Similarly, upon photo-excitation, the
groundstate will be any one of the degenerate maximally spin-polarized
states
\begin{equation}
\label{eq:hole}
h_{m}^\dagger |0\rangle ,
\end{equation}
in which the photo-excited electron fills the vacant spin-$\uparrow$
state, and the hole occupies the lowest Landau level single particle
state with quantum number $m$ ($h_{m}^\dagger$ is the operator that
creates a valence-band hole in this state; we continue to suppress the
subband-label of the hole). We will refer to this state as the
``free-hole state''. This is the simplest ``charged complex'' that can
compete with the excitonic states to be the absolute groundstate of
the system, and therefore to contribute to the low-temperature
photoluminescence spectrum.

We will compare the energy of the free-hole state with that of an
excitonic state in which the valence-band hole forms a $\bbox{P}=0$
exciton with a spin-$\downarrow$ electron a long distance from the
positive quasiparticle. One can convert the free-hole state to this
excitonic state by (1) introducing a widely separated
quasielectron/quasihole pair far from the valence-band hole (at an
energy cost of $Z+B_{SW}$, where $B_{SW}$ is the interaction
contribution to the energy gap of the $\nu=1$ state which we refer to
as the ``binding energy'' of a spin-wave), and (2) binding the
quasielectron to the free valence-band hole (with an energy gain of
$B_X$, which is the binding energy of the exciton). The energy of the
excitonic state is therefore larger than that of the free-hole state
by an amount $Z+B_{SW}-B_X$. For a Zeeman energy $Z$ that is large
compared to the interaction energies, this quantity will be positive,
and the free-hole state will be the lower energy state.  For small
$Z$, as is typically the case experimentally, whether the free-hole or
the exciton state is the lower in energy depends on the relative sizes
of the spin-wave and exciton binding energies.  In
appendix~\ref{app:exciton} it is shown that, within the approximation
of no Landau level mixing for the electrons, these binding energies
may easily be calculated.  The binding energy of the exciton is found
to be independent of the mass of the valence-band hole
\begin{equation}
\label{eq:B_X}
B_X= \sqrt{\frac{\pi}{2}}e^{d^2/2\ell^2}\mbox{erfc}(d/\sqrt{2}\ell)
\frac{e^2}{4\pi\epsilon\epsilon_0\ell}.
\end{equation}
The binding energy of the spin-wave follows from the $d=0$ limit of
Eq.(\ref{eq:B_X})
\begin{equation}
\label{eq:B_SW}
B_{SW}= \sqrt{\frac{\pi}{2}}
\frac{e^2}{4\pi\epsilon\epsilon_0\ell}.
\end{equation}
From these expressions one finds that for any non-zero $d$ the binding
energy of the exciton is {\it less} than that of the spin-wave. The
free-hole state will therefore always be the lower energy state, and,
for $\nu\lesssim 1$, one can expect to see radiative recombination
from the free-hole state rather than from excitonic states.  The form
of the recombination spectrum of the free-hole state is trivial: since
there are no spin-$\downarrow$ electrons present, the hole can only
recombine in the $\sigma_-$ polarization, and will contribute a single
sharp line.  Simple considerations show that the recombination of the
free-hole state occurs at an energy $B_{SW}-B_X$ {\it lower} than the
recombination energy of the $\bbox{P}=0$ excitonic state in this
$\sigma_-$ polarization.

Considerations similar to those we have just presented formed the
basis of the main point of Ref.\onlinecite{muzykantskii}, in which it
was argued that as the filling fraction is swept from $\nu >1$ to
$\nu<1$, the form of the initial state contributing to
photoluminescence undergoes a transition from an excitonic state to a
free-hole state.  As a result of this transition, a red-shift of the
mean position of the photoluminescence line is expected by an amount
$B_{SW}-B_{X}$.  A red-shift consistent with this behaviour has been
observed in very wide quantum well samples\cite{goldbergoverview}. For
narrow quantum wells, and in particular for the experiments of
Ref.~\onlinecite{plentz}, no such red-shift is observed.  For the
parameter-values appropriate to these experiments, one finds
$B_X=8.0\mbox{meV}$ and $B_{SW}=11.1\mbox{meV}$, so the shift in
energy would be $3.1\mbox{meV}$.  This energy difference is likely to
be over-estimated by our model which neglects the finite thickness of
both the electron and hole subband wavefunctions, but even with these
factors included one would expect the energy shift to be above
experimental resolution. That no red-shift is observed seems to
indicate that for this sample there is no change in the nature of the
initial states as the filling fraction sweeps through $\nu=1$.

In the following we show that the absence of a discontinuity in the
form of the photoluminescence can be explained as a result of Landau
level mixing for the {\it electrons}.  For a small spacing $d$, the
corrections due to this Landau level mixing lead to an {\it decrease}
in $B_{SW}-B_{X}$, which may be sufficient to cause this quantity to
change sign and the {\it excitonic} state to become lower in energy
than the free-hole state.  In this case, as the filling fraction of
the sample is swept through $\nu=1$, the low-energy states will remain
well-described by the excitonic states and there will be no
discontinuity in the form of the recombination spectrum.

We will study the effects of Landau level mixing of the electrons by
considering the changes in the binding energies of the exciton and of
the spin-wave to second order in the Coulomb interaction.  These
corrections are of order $(e^2/\epsilon\ell)^2/(\hbar\omega_e)\sim
e^4m_e/\epsilon^2\hbar^2$, and are therefore independent of the
strength of the magnetic field.  

A calculation of the second-order energy correction to the binding
energy of the spin-wave has been reported in Ref.~\onlinecite{sondhi};
it was found that the binding energy {\it decreases} by an amount
\begin{equation}
-\Delta B_{SW} = 0.58\frac{\left(e^2/4\pi\epsilon\epsilon_0\ell\right)^2}
{\hbar\omega_e},
\end{equation}
which is consistent with a recent quantum Monte Carlo evaluation of
this quantity\cite{qmcgap}. For the parameter-values appropriate to
GaAs systems ($m_e=0.067m_0,\epsilon=12.53$), this decrease is
$6.77\mbox{meV}$ and can be significant compared to the overall
spin-wave binding energy neglecting Landau level mixing [this is
$11.1\mbox{meV}$ for the field strength at which the spectra in
Fig.~\ref{fig:experiment}(b) were measured].

In appendix~\ref{app:mixing} we derive the second-order corrections to
the binding energy of the exciton in the presence of the $\nu=1$
groundstate.  The result is
\widetext
\top{-2.8cm}
\begin{equation}
\label{eq:correction}
\Delta B_{X} = +\left[
\sum_{n\neq 0} \frac{\left[{\cal I}_{2n}(d/\ell)\right]^2}
{2^{2n}n(n!)^2}\right]
\frac{\left(e^2/4\pi\epsilon\epsilon_0\ell\right)^2}
{\hbar(\omega_e+\omega_h)} - \left[
\sum_{n\neq 0} \frac{1}{n n! 2^{2n-1}}{\cal I}_{2n-1}(d/\sqrt{2}\ell)
\right] \frac{\left(e^2/4\pi\epsilon\epsilon_0\ell\right)^2}
{\hbar\omega_e},
\end{equation}
\bottom{-2.7cm}
\narrowtext
\noindent
where ${\cal I}_m(z)$ is a function defined in (\ref{eq:ifun}).  The
first term in this expression represents an {\it increase} in the
binding energy, and accounts for the enhanced binding of an exciton in
the absence of the filled Landau level of spin-$\uparrow$ electrons.
The second term is a {\it decrease} in the binding energy, which
effectively arises from the screening of the electron-hole interaction
by the filled Landau level of spin-$\uparrow$ electrons (which becomes
weakly polarizable when Landau level mixing is included).

We evaluate the above correction to the binding energy of the exciton
by performing the sums in Eq.(\ref{eq:correction}) numerically.  For
the parameter-values appropriate to the sample used to measure
Fig.~\ref{fig:experiment}(b), we find that the binding energy of the
exciton decreases, $-\Delta B_{X}=3.24\mbox{meV}$, by an amount that
is significantly less than the decrease in the spin-wave binding
energy due to Landau level mixing ($6.7\mbox{meV}$).  The resulting
net binding energies of the exciton and spin-wave for these
parameter-values are therefore $B_X=4.76\mbox{meV}$ and
$B_{SW}=4.33\mbox{meV}$.  Thus, the first correction arising from
Landau level mixing for the electrons leads to an exciton binding
energy that is {\it larger} than that of the spin-wave, with
$B_X-B_{SW}=0.4\mbox{meV}$.  Since the difference in binding energies
$B_X-B_{SW}$ is positive and larger than the bare Zeeman energy, the
excitonic state remains the groundstate for $\nu\lesssim 1$.
Introducing Landau level mixing for the electrons, we can therefore
account for the observation that there is no discontinuity in the form
of the photoluminescence spectrum at $\nu=1$ in this sample.

For positive $B_X-B_{SW}$, one further expects that, if the free-hole
states were to become populated, their recombination would appear in
the $\sigma_-$ photoluminescence spectrum at an energy {\it higher}
than that of the emission from the $\bbox{P}=0$ excitonic state in
this polarization.  We suggest that the ``B-peak'' appearing in
Fig.~\ref{fig:experiment}(b) could be due to the recombination of such
states.  This peak is consistent with this interpretation, insofar as
it appears only in the $\sigma_-$ polarization and at an energy above
that of the excitonic recombination line.  Since we use a highly
simplified model for the subband wavefunctions and only include Landau
level mixing for the electrons to lowest order, the uncertainties in
the binding energies we calculate are significant.  The close
similarity between our prediction of an energy spacing of 0.4meV and
the observed spacing ($\simeq 0.5\mbox{meV}$) is purely fortuitous,
and cannot be used as a justification for this interpretation of peak
B.  The main problem with this interpretation is that it requires a
metastable population of the free-hole states.  It is possible that
the relaxation rate of free-hole states is sufficiently small that
their radiative recombination occurs before thermal equilibration is
achieved.  In particular, at temperatures less than the bare Zeeman
energy of the electrons $Z$, the density of spin-$\downarrow$
electrons is vanishingly small and the quenching of the free-hole
states may be suppressed (this is consistent with the appearance of
peak B only at very low temperatures, $k_B T\lesssim Z$\cite{plentz}).

In view of the uncertainty in the energy position of peak B and the
requirement of a non-equilibrium population, the assignment of this
peak to the recombination of the free-hole state is rather
speculative.  However, this interpretation may be tested
experimentally.  Ideally, one would study the evolution of the
spectrum as a function of the separation $d$ (which may be controlled
by studying quantum wells of different widths or by the use of front
and back-gates\cite{heimangates}).  As $d$ increases the energy
difference between the recombination from the free-hole and that of
the exciton should {\it decrease} due to the decreasing binding energy
of the exciton.  For sufficiently large $d$, the binding energy of the
exciton $B_X$ will become less than that of the spin-wave $B_{SW}$
plus the Zeeman energy $Z$, and the free-hole state will become the
groundstate configuration for $\nu < 1$.  At this point, one will
recover the behaviour described in Ref.~\onlinecite{muzykantskii} and
observed in wide quantum wells\cite{goldbergoverview}, in which the
photoluminescence line shows a discontinuous red-shift as the filling
fraction is reduced through $\nu=1$.  The transition between these two
regimes occurs at a critical value of the separation $d_c$ which is
defined by the condition that $Z+B_{SW}-B_X=0$ (where $B_{SW}$ and
$B_X$ are the exact binding energies of the spin-wave and exciton,
including all Landau level mixing corrections).  This critical value
may be estimated using Eqs.~(\ref{eq:B_X}--\ref{eq:correction}).  We
find that $d_c$ decreases slowly for samples with increasing carrier
densities, due the reduction of the influence of Landau level mixing
as the magnetic field increases to maintain $\nu=1$.  Using the
parameters appropriate for GaAs ($m_e=0.067, m_h=0.34$,
$\epsilon=12.53$ and an electron $g$-factor of $0.4$), we find
$d_c\simeq 0.5\ell$ at $B=4\mbox{T}$ and $d_c\simeq0.3\ell$ at
$B=10\mbox{T}$.

\subsection{Additional negative charge}
\label{subsec:negative}

Consider a system that, prior to photo-excitation, contains a single
additional negative charge.  For large electron Zeeman energy and in
the absence of disorder the groundstate is one of the maximally
spin-polarized states
\begin{equation}
e_{m\downarrow}^\dagger |0\rangle,
\end{equation}
which are degenerate in $m$.  Upon photo-excitation, the additional
electron must also be spin-$\downarrow$, there being no vacant
spin-$\uparrow$ electron states.  The energy eigenstates of the
resulting maximally-spin-polarized system are determined from the
3-body problem in which two (spin-$\downarrow$) electrons in the
lowest Landau level interact with the valence-band hole.  In this
section we discuss the possibility of a bound-state of all three
particles forming.  It is known that such a bound state does exist for
$d=0$, both when the hole is restricted to the lowest Landau level
($m_h=0$)\cite{rezayi,muzykantskii}, and when the hole mass is
infinite (in which case the system represents the high-field triplet
``$D^-$'' complex\cite{dzyubenko}).  If the energy of this complex is
sufficiently less than that of a widely-separated exciton and
quasiparticle, one can expect a new feature in the photoluminescence
spectrum to appear when $\nu\gtrsim 1$.

We have calculated the binding energy of the exciton to the second
electron numerically for arbitrary $m_h$ and $d$.  We work in the
spherical geometry with system sizes of up to 51 single-particle
states in the lowest Landau level, and extrapolate the binding energy
to the infinite-size limit by using quadratic regression in one over
the number of single-particle states.  To account for Landau level
mixing of the hole, we retain the first five hole Landau levels; this
is sufficient to reproduce the binding energy for even the case
$m_h=\infty, d=0$\cite{dzyubenko} (in which Landau level mixing will
be most important) to an accuracy of $5\%$.  For all values of $d$ for
which the results of our finite-size calculations are reliable
($d\lesssim 2\ell$), we find that the negatively-charged exciton is
bound (the total angular momentum of the groundstate changes as $d$
increases, the first change occurring when this spacing is larger than
about one magnetic length).  For the parameter-values appropriate to
the experiments of Fig.~\ref{fig:experiment}(b), we find a binding
energy of $0.086 e^2/4\pi\epsilon\epsilon_0\ell = 8.8\mbox{K} =
0.76\mbox{meV}$.  This binding energy is large compared to the typical
thermal energy, so one could expect these states to provide an
important contribution to photoluminescence for $\nu\gtrsim 1$.

Since the negatively-charged initial state contains both
spin-$\uparrow$ and spin-$\downarrow$ electrons, one might expect that
this state would radiatively decay in both polarizations.  However,
for all finite values of $m_h$ and $d$, our numerical studies show
that the transition rate in the $\sigma_+$ polarization is identically
zero.  This transition is forbidden by the selection rule arising from
the conservation of total angular momentum by the operator $L_+$,
since our calculations show that the total angular momentum of the
initial state differs from that of the available final state (a single
spin-$\downarrow$ electron).

In the $\sigma_-$ polarization, there is a significant transition rate
for all values of the model parameters.  In the final state there are
two spin-$\downarrow$ electrons and a single spin-hole, appearing as a
result of the recombination of one spin-$\uparrow$ electron with the
hole.  The groundstate of this three-body system is a small charged
spin-texture, in which all three particles are bound closely
together\cite{rezayi,sondhi,macdonaldskyrmion}. To higher energy there
is a continuum of excited states, representing the unbound motion of a
spin-wave in the presence of the additional electron.  It appears from
our numerical calculations, and from an analytic treatment of
particles with hard-core repulsion\cite{macdonaldskyrmion}, that there
is only one bound state, so the final-state energy spectrum consists
of the charged spin-texture state and the spin-wave continuum,
separated by a single energy gap.  This energy determines the
threshold value of the Zeeman energy below which the first
spin-texture becomes lower in energy than the spin-polarized
quasiparticle\cite{sondhi}.

\begin{figure}
\vskip-0.2cm
\inseps{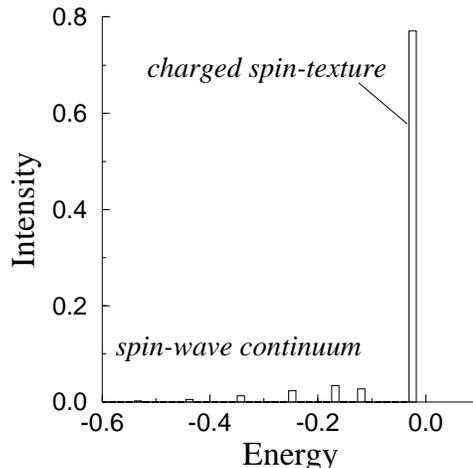}{0.55}
\caption{$\sigma_-$ recombination spectrum  of the
negatively-charged complex, calculated for a sphere with 51
single-particle states in the lowest Landau level and with the
parameter-values $\hbar\omega_h/(e^2/4\pi\epsilon\epsilon_0\ell)=0.15,
d/\ell=6/13$.  The recombination energy is measured in units of
$e^2/4\pi\epsilon\epsilon_0\ell$ relative to that of the $P=0$ exciton
recombination.}
\label{fig:xminusspectrum}
\end{figure}

The recombination spectrum, calculated numerically for the
parameter-values appropriate to the experiments of
Ref.~\onlinecite{plentz}, is shown in Fig.~\ref{fig:xminusspectrum}.
The main peak contains $88\%$ of the total intensity, and is due to
the recombination into the groundstate: the charged spin-texture.  The
remaining $12\%$ is into the unbound spin-wave states.  The finite
size of our system causes this part of the spectrum to be discrete.
In the limit of infinite systems sizes, this will become continuous
and only the gap separating the spin-wave continuum from the charged
spin-texture complex will remain.  We therefore find that the
recombination spectrum in this polarization provides a direct
measurement of the formation energy of the smallest charged
spin-texture.  The observation of such structure in photoluminescence
would be of great interest.

The relative intensities in the charged-spin texture peak and the
spin-wave continuum vary with the parameters $m_h$ and $d$.  If the
hole is restricted to states in the lowest Landau level ($m_h=0$), it
is found that recombination occurs almost exclusively into the
charged-spin texture peak for all values of $d$\cite{muzykantskii}
(provided $d$ is not so large that the total angular momentum of the
initial groundstate changes). In that case, there is vanishingly small
intensity in the spin-wave continuum, so the energy gap is not
pronounced.  It is only when one introduces a finite hole-mass that
appreciable intensity is found in the spin-wave continuum and the gap
can be observed.

Despite the fact that our model predicts this negatively-charged
initial state to be bound for the parameter-values appropriate for the
sample used in Ref.~\onlinecite{plentz}, there is no feature in the
observed spectra [Fig.~\ref{fig:experiment}(b)] which is clearly
associated with the recombination of such a state.  It is possible
that such recombination is masked by the low-energy tail of the
exciton recombination but is responsible for the shoulder observed in
the $\sigma_-$ spectrum of Fig.~\ref{fig:experiment}(b).  It is also
possible that the negatively-charged state is not bound in practice,
due to factors left out of the above calculation of the binding
energy.  One can expect a reduction in the binding energy of this
state due to the finite thicknesses of the subband wavefunctions,
Landau level mixing for the electrons, and the screening due to
spin-depolarization (e.g. the binding of the exciton to the
quasiparticle is likely to be reduced if the Zeeman energy is
sufficiently small that the lowest-energy quasiparticle is a charged
spin-texture).  The negatively-charged state will be most stable in
samples with small values of $d/\ell$, for which the binding energy we
calculate is large, and with high-densities, such that $\nu=1$ occurs
at larger magnetic field and Landau level-mixing and
spin-depolarization effects are less important.

\section{Summary}
\label{sec:summary}

We have studied a model for the low-temperature photoluminescence of
high-mobility quantum wells in the vicinity of the $\nu=1$ quantum
Hall state.  Within this model, we discussed the polarization-resolved
photoluminescence spectra, taking account of a separation $d$ between
the planes in which the electrons and hole move, and Landau level
coupling for the valence-band hole described by a finite effective
mass $m_h$.

The low-energy states at $\nu=1$ are ``excitonic states'' if the
electron Zeeman energy $Z$ is large, or even for vanishing $Z$ if $d$
is not too large ($d<1.3\ell$).  These are states in which the
spin-$\uparrow$ lowest Landau level is filled, and the valence-band
hole binds with a single spin-$\downarrow$ electron to form an
exciton.  The radiative recombination processes of these states are
quite different for the two polarization channels.  In the $\sigma_+$
polarization, the valence-band hole recombines with the
spin-$\downarrow$ electron to which it is bound to leave the $\nu=1$
groundstate: the resulting recombination line is narrow (limited only
by disorder).  In the $\sigma_-$ polarization, the valence-band hole
recombines with one of the spin-$\uparrow$ electrons, and a spin-wave
excitation is left in the final state. The shake-up of spin-waves
causes the recombination line in this polarization to be broadened to
low energy. We argue that the observations of Ref.~\onlinecite{plentz}
demonstrate the recombination of excitonic states.  The long-range
disorder arising from the remote ionized donors leads to strong
scattering of the excitonic states. A simple model for the
recombination of the exciton in this disordered potential leads to
line shapes that compare favourably with the experimental
observations.

We addressed the behaviour at filling fractions slightly away from
$\nu=1$, by considering the photoluminescence of a system containing a
single additional positive ($\nu\lesssim1$), or negative
($\nu\gtrsim1$) quasiparticle.  We compared the energies of the
excitonic states with other ``charged'' initial states that can form
in these cases.  For $\nu\lesssim 1$, we showed that, as a result of
Landau level mixing for the electrons, the excitonic state is the
groundstate for small $d$; for large $d$, a ``free-hole state'' is
lower in energy.  The observation of peak B in the experimental
spectra reported in Ref.~\onlinecite{plentz} is consistent with a
metastable population of the free-hole state; further experiments are
required to justify this assignment.  For $\nu\gtrsim 1$, a
negatively-charged state, in which the valence-band hole binds with
two spin-$\downarrow$ electrons, is lower in energy than the excitonic
state.  The recombination spectrum of this state in the $\sigma_-$
polarization contains information directly related to the formation
energy of the smallest charged spin-texture of this system.  No clear
evidence for this negatively-charged state is observed in the present
experiments.  This state will be more stable and its recombination may
be more clearly observable in higher-density samples with small values
of $d/\ell$.

Finally, we note that similar considerations can be applied to
photoluminescence close to higher integer filling fractions $\nu=i$.
Many of the qualitative features of our model at $\nu=1$ appear also
at these filling fractions. In particular, in the strong-field limit,
there are low-energy initial states similar to the excitonic, and
positively- and negatively-charged initial states described above, but
with the electrons forming the excitonic and negatively-charged
complexes now lying in a high Landau level.  Discontinuities similar
to those discussed in Sec.\ref{subsec:positive} can arise as the
filling fraction is swept from $\nu\gtrsim i$ to $\nu\lesssim i$, as a
result of a change in the groundstate from an excitonic (or
negatively-charged) state to a positively-charged state.  Although we
have not calculated detailed spectra in these cases, it is clear that
the recombination of excitonic initial states will again lead to
asymmetric line shapes, due either to the shake-up of spin-waves at
odd filling fractions $\nu=2n+1$, or to the shake-up of
magnetoplasmons and magnetoplasmons combined with spin-flips at all
integer filling fractions $\nu=i>1$.

\acknowledgements{We are grateful to Bertrand I. Halperin for
advice on the theoretical aspects of this work and to Don~Heiman and
Flavio~Plentz for discussions concerning their experiments and for
providing us with experimental data and unpublished information
regarding their samples.  This work was partially supported by the
NATO Science Fellowship Programme, by the Harvard Society of Fellows
and by NSF grant DMR 94-16910.}

\widetext
\top{-2.8cm}

\begin{appendix}

\section{Exciton States in Strong Magnetic Field}

\label{app:exciton}

\subsection{Free Exciton States}

\label{appsec:freeexciton}

We study a Hamiltonian of the form
\begin{equation} 
\label{eq:fullhamiltonian} 
H = \frac{(
\bbox{p}_e + e\bbox{A}_e)^2}{2m_e}
 + \frac{( \bbox{p}_h - e\bbox{A}_h)^2}{2m_h} + V^{eh} (\bbox{r}_e -
\bbox{r}_h),
\end{equation} 
where the subscripts $e$ and $h$ refer to the electron and hole
co-ordinates, and we will choose to work in the symmetric gauge,
$\bbox{A}(\bbox{r})= \bbox{B}\times\bbox{r}/2$.  This two-body problem
was greatly simplified by Gorkov and Dzyaloshinskii\cite{gorkov} who
showed that the energy eigenstates may be classified by the
eigenvalues of a conserved momentum
\begin{equation}
\label{eq:ganddmomentum}
{\bbox{P}} \equiv {\bbox{p}}_e + {\bbox{p}}_h - \frac{e}{2}
\bbox{B}\times ({\bbox{r}}_e - {\bbox{r}}_h) .
\end{equation}
Replacing this operator by its eigenvalue, one finds that the energy
eigenstates are determined by a one-body Schr\"odinger equation for
the relative co-ordinate $\bbox{r}\equiv \bbox{r}_e - \bbox{r}_h$ with
the Hamiltonian
\begin{equation}
\label{eq:relativehamiltonian}
H_P = -\frac{\hbar^2}{2\mu} \nabla^2 + 
\frac{i e\hbar B\gamma }{2\mu} \bbox{r}\times \nabla
+  \frac{e^2B^2}{8\mu}\bbox{r}^2 
+ V^{eh}(\bbox{r})
+ \frac{\bbox{P}^2}{2M} +
\frac{eB}{M}(\bbox{P}\times\hat{\bbox{z}}).\bbox{r},
\end{equation}
where $M\equiv m_e+m_h$, $\mu\equiv m_em_h/(m_e+m_h)$, and $\gamma
\equiv (m_h-m_e)/(m_e+m_h)$.
As a result of these transformations, the overall eigenfunctions 
of Eq.(\ref{eq:fullhamiltonian})
take the form
\begin{equation}
\Psi(\bbox{r}_e,\bbox{r}_h) = \frac{1}{\sqrt{\Omega}}
e^{i\bbox{P}.(\bbox{r}_e+\bbox{r}_h)/2\hbar}
e^{i\bbox{r}_e\times\bbox{r}_h/2\ell^2}
e^{-i\gamma\bbox{P}.(\bbox{r}_e-\bbox{r}_h)/2\hbar}
\psi_P(\bbox{r}_e-\bbox{r}_h),
\end{equation}
\bottom{-2.7cm}
\narrowtext
\noindent
where $\psi_P(\bbox{r})$ is the solution of
Eq.(\ref{eq:relativehamiltonian}), and the state is normalized to an
area $\Omega$.

The eigenstates of this Hamiltonian were discussed by Lerner and
Lozovik\cite{lernloz} in the limit $m_e, m_h\rightarrow 0$ in which
Landau level mixing for both the electron and hole may be neglected.
We do not follow this approach, as we are interested in cases for
which the hole mass is finite.  Rather, we will study the limit
$m_e\rightarrow 0$, $m_h=\mbox{finite}$, in which there is no Landau
level mixing for the electron, but there may be for the hole.

We begin by finding the eigenstates of (\ref{eq:relativehamiltonian})
for $\bbox{P}=0$.  Initially we neglect the electron-hole interaction,
and solve for the eigenstates of the kinetic energy operator.  These
are the same as the symmetric gauge Landau level states $|n,m\rangle$
characterized by the Landau level index $n$ and angular momentum $m$.
The energies are\cite{lernloz}
\begin{eqnarray}
\nonumber
E_{n,m} &  = & \hbar \omega_e \left[n + (|m|-m)/2 +1/2\right] + \\ 
& & \hbar \omega_h \left[n + (|m|+m)/2 +1/2\right],
\label{eq:enm}
\end{eqnarray}
where $\omega_{e,h} \equiv eB/m_{e,h}$.  The state $|n,m\rangle$ is
therefore a state in which the electron has a Landau level index
$n_e=n + (|m|-m)/2$ and the hole an index $n_h=n + (|m|+m)/2$.  We now
introduce the electron-hole interaction.  The approximation that we
make is that the cyclotron energy of the electron, $\hbar
\omega_e$, is large compared to the interaction energy, such that
coupling between states with different $n_e$ may be neglected.  In
particular, for this work we focus on the states $|n=0,m\rangle$ in
which the electron is in the lowest Landau level and the hole is in
the $m^{\mbox{\tiny th}}$ Landau level.  Within this approximation,
the states $|n=0,m\rangle$ are eigenstates of the electron-hole
interaction, $V^{eh}(\bbox{r})$, since this potential conserves the
angular momentum $m$.  The energies of these states are found from the
expectation value of the interaction
\begin{equation}
\label{eq:exactenergies}
E_{0,m}(\bbox{P}=0) = \frac{1}{2}\hbar \omega_e + 
\hbar \omega_h (m+1/2)
+ V^{eh}_m
\end{equation}
where we have defined $V^{eh}_m\equiv\langle 0,m|V^{eh}|0,m\rangle$.
These are the energies of the $\bbox{P}=0$ exciton states for which
the electron is in the lowest Landau level.  We will concentrate on
the lowest-energy state, $m=0$, for which the binding energy is
$-V^{eh}_0$.  This binding energy may be calculated exactly for the
interaction
\begin{equation}
\label{eq:veh}
V^{eh}(\bbox{r}) = - \frac{e^2}{4\pi\epsilon\epsilon_0}
\frac{1}{\sqrt{|\bbox{r}|^2+d^2}} ,
\end{equation}
which represents electrons and holes moving in planes separated by
$d$. We find\cite{cooperwc}
\begin{equation}
\label{eq:veh0}
- V^{eh}_0 =  \frac{e^2}{4\pi\epsilon\epsilon_0 \ell} 
\sqrt{\pi/2} \; e^{d^2/2\ell^2}
\mbox{erfc}\left(d/\sqrt{2}\ell\right) .
\end{equation}
As $d/\ell\rightarrow 0$ this expression recovers the binding energy
of the exciton calculated by Lerner and Lozovik\cite{lernloz} for the
case in which the electron and hole move in a single plane. Note that
both the wavefunction of the zero-momentum exciton state and the above
binding energy are independent of the value of the hole effective
mass.

The effective mass of each of these exciton states may be found
exactly by treating the momentum-dependent terms in
(\ref{eq:relativehamiltonian}) within a perturbation expansion. We
concentrate only on the lowest-energy exciton state $n=0, m=0$.  To
second order in the momentum, the change in energy is
\begin{eqnarray}
\Delta E_{0,0}(\bbox{P}) & = & \frac{\bbox{P}^2}{2M} + \frac{e^2B^2}{M^2}\sum_{m
\neq 0}
\frac{ | \langle 0,m|\bbox{r}|0,0\rangle
.\bbox{P}\times\hat{\bbox{z}}|^2}{V^{eh}_0 - (m\hbar\omega_h+
V^{eh}_m)} 
\nonumber
\\
\label{eq:fulleffectivemass}
 & = & \frac{\bbox{P}^2}{2m_h}
\left[1+\frac{\hbar\omega_h}{V^{eh}_1-V^{eh}_0}\right]^{-1}.
\end{eqnarray}
In the last line we have set $M=m_h$, since our analysis is strictly
correct only in the limit $m_e\rightarrow 0$.  The effective mass of
the exciton is therefore
\widetext
\top{-2.8cm}
\begin{eqnarray}
M_X & = & m_h+\frac{\hbar^2}{\ell^2}\frac{1}{V^{eh}_1-V^{eh}_0}
\nonumber
\\
 & = & m_h+\frac{4\pi\epsilon\epsilon_0\hbar^2}{e^2\ell}\left[
\sqrt{\frac{\pi}{8}}\exp{(d^2/2\ell^2)}(1+d^2/\ell^2)\mbox{erfc}(d/\sqrt{2}\ell)
-d/2\ell\right]^{-1}.
\label{eq:exacteffectivemass}
\end{eqnarray}
The first-order corrections to the internal wavefunction, lead to the
overall wavefunction of the exciton groundstate
($\hbar=\ell=1$)
\begin{eqnarray}
\Psi_{\bbox{P}}(\bbox{r}_e,\bbox{r}_h) & \simeq & \frac{1}{\sqrt{2\pi \Omega}}
e^{i\bbox{P}.\bbox{r}_h} e^{i\bbox{r}_e\times
\bbox{r}_h.\hat{\bbox{z}}/2} e^{-(\bbox{r}_e-\bbox{r}_h)^2/4} \left[ 1
+ \frac{1}{2(1+\lambda)} (i\bbox{P} -
\bbox{z}\times\bbox{P}).(\bbox{r}_e-\bbox{r}_h)\right] 
\nonumber
\\
\label{eq:wavefunction}
& \simeq &
\frac{1}{\sqrt{2\pi \Omega}}
e^{i\bbox{P}.[\bbox{r}_e+(1+2\lambda)\bbox{r}_h]/(2+2\lambda)}
e^{i\bbox{r}_e\times \bbox{r}_h/2}
e^{-[\bbox{r}_e-\bbox{r}_h-\bbox{r}_P/(1+\lambda)]^2/4} ,
\end{eqnarray}
\bottom{-2.7cm}
\narrowtext
\noindent
where $\bbox{r}_P\equiv \hat{\bbox{z}}\times\bbox{P} \ell^2/\hbar$,
and we have set $m_e=0$.
The parameter $\lambda$ is defined to be
\begin{equation}
\label{eq:lambda}
\lambda \equiv (V^{eh}_1-V^{eh}_0)/\hbar\omega_h,
\end{equation}
which is a measure of the extent of Landau level mixing for the hole.

The above perturbative results for the dispersion relation and
wavefunctions are accurate for $|\bbox{P}|\ll (1+\lambda)\hbar/\ell$.
In the last line of Eq.(\ref{eq:wavefunction}) we have introduced
exponential functions which reproduce the first-order corrections in
the momentum.  These are chosen such that the expression correctly
reproduces the {\it exact} wavefunction in the limit of no Landau
level coupling\cite{lernloz}, $\lambda\rightarrow 0$, and much Landau
level coupling, $\lambda\rightarrow \infty$, for the hole.

In the limit of no Landau level mixing for the hole,
$\lambda\rightarrow 0 $, the exciton effective mass is due only to the
interaction, and we can recover the mass derived by Lerner and Lozovik
for an electron and hole moving in the same plane and neglecting all
Landau level coupling by setting $d=0$:
$M_X=\sqrt{8/\pi}(4\pi\epsilon\epsilon_0\hbar^2)/(e^2\ell)$.  In this
limit, the above expressions for the binding energy, effective mass
and wavefunction of the exciton also describe the properties of
spin-waves at $\nu=1$ when Landau level mixing is
neglected\cite{bychkov,kallinhalperin1}. In this case, the electron of
the exciton represents a spin-$\downarrow$ electron in the lowest
Landau level, and the hole a missing spin-$\uparrow$ electron in an
otherwise filled band of spin-$\uparrow$ electrons in the lowest
Landau level.  The binding energy of the exciton determines the
spin-wave bandwidth, and the inverse effective mass of the exciton
gives the spin-stiffness.

\subsection{Exciton states in a smooth external potential}

\label{subapp:externalpotential}

In this section we show how one may obtain an effective dynamics for
the motion of the exciton in an external potential that is
sufficiently weak and long-ranged.  Our procedure is analogous to the
Born-Oppenheimer approximation in the theory of molecular dynamics.
In that case, a simplification arises due to the separation of
timescales between the fast electronic motion and the slow atomic
motion.  In the present case, the relative motion of the electron-hole
pair is treated as ``fast'', and the centre-of-mass motion of the
exciton is assumed to be much slower.

To be specific, we introduce the potentials $V^e(\bbox{r}_e)$ and
$V^h(\bbox{r}_h)$ in the Hamiltonian (\ref{eq:relativehamiltonian}).
The momentum $\bbox{P}$ is no longer conserved.  However, within the
spirit of the Born-Oppenheimer approximation, we imagine that the slow
co-ordinates, the centre-of-mass momentum $\bbox{P}$ and position
$\bbox{R}\equiv (m_e\bbox{r}_e+ m_h\bbox{r}_h)/(m_e+m_h)$ (which are
canonically conjugate), are stationary, and solve for the groundstate
of the internal motion.  The resulting energy function serves as an
effective Hamiltonian for the centre-of-mass motion.  This procedure
is appropriate provided the energy separation between the resulting
centre-of-mass states is much smaller than the energy spacing between
the internal states of the exciton.

To simplify this procedure, we expand the external potentials to first
order in the relative co-ordinate
\widetext
\top{-2.8cm}
\begin{equation}
V^e(\bbox{r}_e)+V^h(\bbox{r}_h) = V^e(\bbox{R}) + V^h(\bbox{R}) +
\bbox{r}.\left[\eta_h\nabla V^e(\bbox{R}) -
\eta_e\nabla V^h(\bbox{R})\right], 
\end{equation}
where $\eta_{e,h}\equiv m_{e,h}/(m_e+m_h)$, and it is to be understood
that the gradient operators apply in the plane of motion. This
approximation requires the potentials to be smooth on the lengthscale
of the exciton size, $\ell$.  Since the potentials now couple to the
internal motion through a term proportional to $\bbox{r}$, we can
calculate the second-order energy shift using the same perturbative
approach we used above.  Combining this with the kinetic energy
(\ref{eq:fulleffectivemass}) we obtain
\begin{equation}
\Delta E_{0,0}   =  \frac{\bbox{P}^2}{2M_x}
+ V^e(\bbox{R}) + V^h(\bbox{R}) + \frac{1}{1+\lambda}\frac{\ell^2
\bbox{P}}{\hbar}.  (\eta_h\nabla V^e - \eta_e\nabla V^h)\times
\hat{\bbox{z}} 
- \frac{\lambda}{1+\lambda}\frac{\ell^2}{2}\frac{\left|
\eta_h\nabla V^e - \eta_e
\nabla V^h\right|^2}{V^{eh}_1-V^{eh}_0},
\end{equation}
\bottom{-2.7cm}
\narrowtext
\noindent
where the effective mass $M_X$ is that defined in
(\ref{eq:exacteffectivemass}) and $\lambda$ is defined in
(\ref{eq:lambda}).  It is to be understood that the term that is
linear in momentum should be symmetrized with respect to momentum and
position operators, such that the Hamiltonian is Hermitian.

This expression represents the effective Hamiltonian for the
centre-of-mass motion of the exciton.  The approximations used to
derive this were (1) the centre-of-mass motion is ``slow'' compared to
the internal motion, and (2) the external potentials are smooth on the
scale of the magnetic length.

\section{Calculation of the Disorder-Dominated Spectral Line Shape}
\label{app:disorder}

In this appendix we provide some details of the steps required to
calculate line shapes for the exciton recombination (in both
polarizations) in the presence of long-range disorder, within the
simple model outlined in Sec.~\ref{subsec:disorder}.  This model
averages over the recombination spectra of exciton groundstates in all
potential minimum of $V^X(\bbox{r})$, expanding the potential in the
vicinity of each minimum to harmonic order and assigning equal weight
to each (harmonic-oscillator) groundstate. One therefore must know (1)
the spectrum of radiation emitted from the groundstate in a given
potential minimum (characterized by the potential and its two
principal curvatures), and (2) the distribution of these minima for
the form of disorder in which we are interested.

We begin by calculating the recombination spectra for the exciton
groundstate in a potential minimum described by
\begin{equation}
V(\bbox{r}) = V + \frac{1}{2} V_{\xi\xi} \xi^2 +
\frac{1}{2}  V_{\eta\eta} \eta^2 ,
\end{equation}
where $\xi$ and $\eta$ are the distances from the centre of the
minimum along the principal axes (for convenience of notation, in this
section we omit the $X$ superscript on the exciton potential energy).
The potential minimum is characterized by the three parameters $\{V,
V_{\xi\xi}, V_{\eta\eta}\}$.  For an exciton with a parabolic
dispersion, the Hamiltonian factorizes into two 1-D simple harmonic
oscillators and the spectrum and wavefunctions may be easily found.
The groundstate energy is
\begin{equation}
E_0(V, V_{\xi\xi}, V_{\eta\eta}) = V +
\frac{\hbar}{2}\sqrt{\frac{V_{\xi\xi}}{M_X}} +
\frac{\hbar}{2}\sqrt{\frac{V_{\eta\eta}}{M_X}},
\label{eq:gsenergy}
\end{equation}
where $M_X$ is the exciton effective mass. The groundstate
wavefunction is the product of two gaussian functions of $\xi$ and
$\eta$.

In the $\sigma_+$ polarization, radiative recombination of the exciton
leaves a single final state (the $\nu=1$ groundstate).  The spectrum
of recombination for the exciton groundstate in the potential minimum
with $\{V, V_{\xi\xi}, V_{\eta\eta}\}$ is therefore simply
\begin{equation}
I_+(E; V, V_{\xi\xi}, V_{\eta\eta}) = \delta[E - E_0(V, V_{\xi\xi},
V_{\eta\eta})],
\label{eq:iplus}
\end{equation}
where the energy $E$ of the emitted radiation is measured relative to
the recombination of the free $P=0$ exciton in this polarization.

In the $\sigma_-$ polarization, the exciton annihilates to leave a
spin-wave in the final state. Due to the finite spatial extent of the
initial exciton state, the final spin-wave state is a superposition of
many momentum components and therefore of many energy eigenstates
(since the spin-waves do not feel the disorder potential they behave
as free particles).  The transition therefore has a finite width in
energy, of approximately $\hbar^2/M_{SW} R^2$ where $R$ is a measure
of the spatial extent of the groundstate wavefunction and $M_{SW}$ is
the spin-wave effective mass.  In detail, the spectrum for
recombination of the exciton groundstate $|\psi(V, V_{\xi\xi},
V_{\eta\eta})\rangle$ is
\widetext
\vskip1cm
\begin{eqnarray}
I_-(E; V, V_{\xi\xi}, V_{\eta\eta}) & \equiv &
\sum_{\bbox{P}} | _{SW}\langle \bbox{P} | \hat{L}_-| \psi \rangle|^2
\delta[E-(E_0-\bbox{P}^2/2M_{SW})] \\
& = & \sum_{\bbox{P}} |\langle \bbox{P}| \psi\rangle|^2
\delta[E-(E_0-\bbox{P}^2/2M_{SW})],
\end{eqnarray}
where we have made use of the matrix
element~(\ref{eq:xswmatrixelementfinal}) to relate $I_-(E)$ to the
momentum distribution of the exciton wavefunction,
$\langle\bbox{P}|\psi\rangle$. Again, the emission energy is measured
relative to that of a free $P=0$ exciton in this polarization. Using
the explicit form of the groundstate harmonic oscillator wavefunction,
this line shape is found to be
\begin{equation}
I_-(E; V, V_{\xi\xi}, V_{\eta\eta})   = 
\frac{2M_{SW}/M_X}{\sqrt{E_{\xi\xi}E_{\eta\eta}}}
\exp\left[(E-E_0)\frac{M_{SW}}{M_X}(E_{\xi\xi}^{-1}+E_{\eta\eta}^{-1}
)\right] \times  
I_0\left[(E-E_0)\frac{M_{SW}}{M_X}(E_{\xi\xi}^{-1}-E_{\eta\eta}^{-1})\right]
\Theta(E_0-E),
\label{eq:iminus}
\end{equation}
where $E_{ii}\equiv \hbar\sqrt{V_{ii}/M_X}$ for $i=\{\xi,\eta\}$,
$E_0$ is the energy~(\ref{eq:gsenergy}), $I_0$ is an imaginary Bessel
function, and $\Theta(z)$ is the Heaviside step function.

We now know the line shapes (\ref{eq:iplus},\ref{eq:iminus}) for the
two polarizations of the emission from the groundstate in the
potential minimum $\{V, V_{\xi\xi}, V_{\eta\eta}\}$.  To compute the
overall spectrum, it is also necessary to know the relative densities
of minima with the characteristics $\{V, V_{\xi\xi}, V_{\eta\eta}\}$.
We will call this distribution $P_{min}(V, V_{\xi\xi}, V_{\eta\eta})$.
For typical statistical forms of the disorder potential, there will be
correlations between the variables $V, V_{\xi\xi}$ and $V_{\eta\eta}$
(potential minima with small values of $V$ are likely to have large
positive curvatures etc.).  Remarkably, however, $P_{min}$ can be
calculated {\it exactly}, including all such correlations, for
disorder potentials that are Gaussian-correlated.  We will now outline
the steps leading to this result, following the approach of
Refs.~\onlinecite{yang,wilkinson} where a similar issue is discussed.

We begin with some general definitions that do not depend on the form
of the disorder.  We denote the joint probability distribution (at any
point in the sample) of $V$ and all of its first and second
derivatives with respect to some fixed axes
$\{\hat{\bbox{x}},\hat{\bbox{y}}\}$ by $P(V,V_x,V_y,V_+,V_-,V_{xy})$,
where $V_\pm\equiv (V_{xx}\pm V_{yy})/2$.  The disorder is assumed to
be homogeneous, such that this function is independent of position.
The density of {\it stationary points} at which $V_x=V_y=0$ and for
which the potential and curvatures are $\{V,V_+,V_-,V_{xy}\}$ may be
expressed in terms of this distribution\onlinecite{wilkinson}
\begin{equation}
P_{stat}(V,V_+,V_-,V_{xy}) = (V_+^2 - V_-^2-V_{xy}^2)\;
P(V,0,0,V_+,V_-,V_{xy}).
\label{eq:pstat}
\end{equation}
For our purposes it is more convenient to work in terms of the
principal curvatures $\{V_{\xi\xi},V_{\eta\eta}\}$, which for a
stationary point with curvatures $\{V_+,V_-,V_{xy}\}$ are
\begin{eqnarray}
V_{\xi\xi} & = & V_+ +\sqrt{ V_-^2 + V_{xy}^2},
\\
V_{\eta\eta} & = & V_+ -\sqrt{ V_-^2 + V_{xy}^2},
\end{eqnarray}
with $V_{\xi\xi}\geq V_{\eta\eta}$ chosen.  One can convert the
distribution~(\ref{eq:pstat}) into the distribution of stationary
points at which the potential and {\it principal curvatures} are
$\{V,V_{\xi\xi},V_{\eta\eta}\}$.  Noting that potential minima are
those stationary points for which both curvatures are positive, we
then find
\begin{eqnarray}
P_{min}(V,V_{\xi\xi},V_{\eta\eta}) & \propto &
\Theta(V_{\xi\xi})\Theta(V_{\eta\eta})
\int dV_+\int dV_-\int dV_{xy}\;
(V_+^2-V_-^2-V_{xy}^2)\; P(V,0,0,V_+,V_-,V_{xy}) 
\nonumber
\\ 
& &
\times \delta\left(V_{\xi\xi} - V_+ -
\sqrt{ V_-^2 + V_{xy}^2}\right) \delta\left(V_{\eta\eta} - V_+ +
\sqrt{ V_-^2 + V_{xy}^2}\right).
\label{eq:pmingeneral}
\end{eqnarray}

To proceed further, we must determine the function
$P(V,V_x,V_y,V_+,V_-,V_{xy})$, which contains all of the relevant
information on the disordered potential.  At this point we specialize
the discussion to disorder potentials which are Gaussian random
functions with zero mean (we choose the zero of energy such that the
average disorder potential vanishes).  In this case, the explicit form
of the distribution $P(V,V_x,V_y,V_+,V_-,V_{xy})$ may be easily found.
It depends only on the averages of all pairwise products of its
variables. The correlations of the gradients $V_x$ and $V_y$ with all
other variables vanish, as do the correlations of $V_-$ and $V_{xy}$
for spatially isotropic disorder (which we now assume).  One
finds\cite{yang,wilkinson}
\begin{equation}
P(V,0,0,V_+,V_-,V_{xy}) \propto 
\exp\left[-\frac{ \overline{V_+^2}\,V^2 + 
 \overline{V^2}\,V_+^2 - 2 \overline{V V_+}\,V V_+}
{2(\overline{V_+^2}\,\overline{V^2} - \overline{V V_+}^2)}\right]
\exp\left[-\frac{V_-^2}{2\overline{V_-^2}}
- \frac{V_{xy}^2}{2\overline{V_{xy}^2}}\right],
\label{eq:pgaussian}
\end{equation}
where the bars denote disorder averages.  For the disorder potential
arising from the ionized donors located a distance $s$ from the
quantum well and with density correlations described
by~(\ref{eq:donorcorrelation}), these coefficients are
\begin{eqnarray}
\overline{V^2} & = & \left(\frac{e^2}{8\pi\epsilon\epsilon_0 s}\right)^2
2\pi n^* d^2  ,
\label{eq:vsq}\\
\overline{V_+^2} & = & \frac{15}{8}\frac{1}{s^4}\overline{V^2} , \\
\overline{V V_+} & = & -\frac{3}{4}\frac{1}{s^2}\overline{V^2}  ,\\
\overline{V_-^2} =  \overline{V_{xy}^2} & = &
\frac{15}{16}\frac{1}{s^4}\overline{V^2},
\label{eq:vminussq}
\end{eqnarray}
where an effective density $n^*$ is used to take account of donor
correlations\cite{pikusefros}, as discussed in
Sec.~\ref{subsec:disorder}.  For the distribution $P$ described by
Eqs.(\ref{eq:pgaussian})--(\ref{eq:vminussq}), the integrals of
Eq.(\ref{eq:pmingeneral}) may be performed to obtain our final
expression for the distribution of minima of the disorder potential in
which we are interested
\begin{equation}
P_{min}(V,V_{\xi\xi},V_{\eta\eta}) \propto 
\Theta(V_{\xi\xi})\Theta(V_{\eta\eta})
\Delta_-
(\Delta_+^2-\Delta_-^2) \exp\left[-\frac{15 V^2 + 8s^4 \Delta_+^2 + 12
s^2 V
\Delta_+}{21\overline{V^2}}-\frac{8s^4\Delta_-^2}{15\overline{V^2}}\right],
\label{eq:pminfinal}
\end{equation}
which we have simplified by defining $\Delta_\pm \equiv (V_{\xi\xi}\pm
V_{\eta\eta})/2$.  As emphasized in Ref.~\onlinecite{yang}, with a
suitable rescaling of energy and lengthscale, the
distribution~(\ref{eq:pgaussian}) depends on the spatial correlations
of the disorder only through the dimensionless parameter $a\equiv
\overline{V_{xy}^2}\,\overline{V^2}/\overline{V_x^2}^2$ ($=5/3$ for the 
form of disorder we study); this is also true for
$P_{min}(V,V_{\xi\xi},V_{\eta\eta})$.

The spectra arising from the recombination of an exciton in a
potential minimum characterized by $\{V,V_{\xi\xi},V_{\eta\eta}\}$
(\ref{eq:iplus},\ref{eq:iminus}) may be combined with the above
distribution for such minima~(\ref{eq:pminfinal}) to obtain the
spectral line shapes within our model
\begin{equation}
I_\pm (E) = \int dV \int dV_{\xi\xi}\int dV_{\eta\eta}
\; P_{min}(V,V_{\xi\xi},V_{\eta\eta}) \; I_\pm (E;
V,V_{\xi\xi},V_{\eta\eta}).
\label{eq:finalsum}
\end{equation}
\bottom{-2.9cm}
\narrowtext
\noindent
This equation simply expresses the assumption that recombination
occurs with equal probability from exciton groundstates in all
potential minima.  We have not been able to find closed form
expressions for the integrals~(\ref{eq:finalsum}), and have therefore
calculated the line shapes numerically, discretizing the
three-dimensional integral by a lattice with $10^6$ points. The
results are shown in Fig.~\ref{fig:disorderlineshapes} for the
parameter-values appropriate to the conditions under which
Fig.~\ref{fig:experiment}(b) was measured.  We will now briefly
discuss the form of the recombination in each polarization.

In the $\sigma_+$ polarization, the line shape depends on a single
dimensionless parameter: the ratio of the typical exciton kinetic
energy to the typical potential energy fluctuation, $\alpha\equiv
(\hbar^2/M_X s^2) /\sqrt{\overline{V^2}}$.  For the parameter-values
we use to compare with experiment this ratio is rather small,
$\alpha=0.25$.  If $\alpha$ were to be zero, there would be no kinetic
energy contribution to the exciton energy, and the spectrum would
simply measure the heights of the minima of the potential. In this
limit our model for the line shape in this polarization reduces to
that proposed in Ref.\onlinecite{yang} for empty quantum wells (with
$a=5/3$, as is appropriate for the form of disorder we consider).  For
non-zero $\alpha$, the energies of all states are increased due to the
non-zero kinetic energy of the exciton, by an amount that differs for
potential minima with differing principal curvatures.  Strictly
speaking, consistency of our model requires that the kinetic energy of
the exciton should always be small compared to the typical potential
fluctuation such that the harmonic approximation is valid; this is
equivalent to the requirement that $\alpha$ be small.

In the $\sigma_-$ polarization, the line shape depends both on
$\alpha$ and on a second dimensionless parameter, $\beta\equiv
M_X/M_{SW}$, which is a measure of the spin-wave stiffness.  For
$\beta=0$, the spin-wave kinetic energy is negligible, and the
recombination spectrum of each exciton state is a sharp line at the
initial exciton energy; the line shape in this polarization becomes
identical to that in the $\sigma_+$ polarization.  For non-zero
$\beta$ each exciton transition is broadened to low energy due to the
shake-up of high-momentum, and hence high-energy, spin waves. The
extent of this broadening depends on the size of the initial exciton
wavefunction and therefore on the principal curvatures of the
potential minimum.  For the parameters-values we use for experimental
comparisons $\beta=6.2$ is large and this broadening is significant.

\section{Correction to the Excitonic Binding Energy due to Landau
level mixing for the Electrons}
\label{app:mixing}

To determine the lowest-order corrections to the binding energy of the
excitonic state, we explicitly calculate the changes in energy of (1)
a zero momentum exciton formed from a spin-$\downarrow$ electron and a
valence-band hole, (2) a single spin-$\downarrow$ electron, and (3) a
single valence-band hole, each in the presence of the filled Landau
level of spin-$\uparrow$ electrons, and (4) the filled Landau level
itself.  In this section, we find it convenient to work in the Landau
gauge $\bbox{A}(\bbox{r}) = B x\hat{\bbox{y}}$, for which the single
particle states for electrons and holes, $\langle \bbox{r}_e
|nk\rangle$ and $\langle nk|\bbox{r}_h\rangle$, are described by the
Landau level index $n$ and a wavevector $k$.  In the absence of Landau
level mixing, the wavefunctions of each of the above states are
\begin{eqnarray}
\Phi_X & = &
\frac{1}{\sqrt{N}}\sum_k e_{k\downarrow}^\dagger h_k^\dagger |0\rangle  ,\\
\Phi_e & = & e_{0\downarrow}^\dagger |0\rangle ,\\
\Phi_h & = & h_{0}^\dagger |0\rangle ,\\
\Phi_0 & = & |0\rangle ,
\end{eqnarray}
where $e_{k}^\dagger$ and $h_{k}^\dagger$ create electrons and holes
in the lowest Landau level states with momentum $k$, and $N=n_0\Omega$
is the number of single particle states in this Landau level.  Note
that these energy eigenstates are independent of all model parameters
($d$, $m_h$).  This is of course true for the cases of the $\Phi_0$,
$\Phi_e$ and $\Phi_h$, and is shown in appendix~\ref{app:exciton} for
the zero-momentum exciton state.

The lowest-order corrections to the energy of the states $\Phi_\alpha$
($\alpha=0,e,h,X$) is
\begin{equation}
\Delta E_\alpha = - \sum_{\Phi_f\neq\Phi_\alpha} 
\frac{|\langle \Phi_f | H | \Phi_\alpha \rangle|^2}{E_f-E_\alpha}
\end{equation}
where $\{\Phi_f,E_f\}$ are a complete set of energy eigenstates and
eigenvalues the full Hamiltonian, $H$, which includes the kinetic
energy of electrons and holes and all interactions $V=V^{ee}+V^{eh}$.
The only non-zero matrix elements involve states $\Phi_f$ in which at
least one electron has a non-zero Landau level index, so to obtain the
energy shift that is correct to order $V^2/\hbar\omega_e$, it is
sufficient to retain only the kinetic energy contribution to the
energies $E_f$ and $E_\alpha$.

By explicit summation over all final states we obtain expressions for
the changes in the exciton, electron and hole energies relative to
that of the filled Landau level, $\Delta E_\alpha-\Delta E_0$, in
terms of the matrix elements of the electron-electron and
electron-hole interactions,
\widetext
\top{-2.8cm}
\begin{eqnarray}
V^{ee}_{n_1k_1,n_2k_2,n_3k_3,n_4k_4} & \equiv &
\int \int d^2\bbox{r}d^2\bbox{r^\prime}
\langle n_1 k_1| \bbox{r}\rangle 
\langle n_2 k_2| \bbox{r^\prime}\rangle 
V^{ee}(\bbox{r}-\bbox{r^\prime})
\langle \bbox{r}| n_4 k_4\rangle 
\langle \bbox{r^\prime}|n_3 k_3\rangle , \\
V^{eh}_{n_1k_1,n_2k_2,n_3k_3,n_4k_4} & \equiv &
\int \int d^2\bbox{r}d^2\bbox{r^\prime}
\langle n_1 k_1| \bbox{r}\rangle 
\langle \bbox{r^\prime}|n_2 k_2\rangle
V^{eh}(\bbox{r}-\bbox{r^\prime})
\langle \bbox{r}| n_4 k_4\rangle 
\langle n_3 k_3| \bbox{r^\prime}\rangle.
\end{eqnarray}
Making use of the invariance of these coefficients under a uniform
displacement of all momenta, the change in the binding energy of the
exciton
\begin{equation}
\Delta B_X \equiv  - ( \Delta E_X-\Delta E_0) + (
\Delta E_e+\Delta E_h-2\Delta E_0) ,
\end{equation}
is found to be
\begin{equation}
\Delta B_X  =  
\sum_{n_e\neq 0, n_h\neq 0, k, k_e, k_h} 
\frac{V^{eh}_{n_ek_e,n_hk_h,00,00}V^{eh*}_{n_e(k_e+k),n_h(k_h+k),00,00}}
{n_e\hbar\omega_e+ n_h\hbar\omega_h}
 + \sum_{n\neq 0, k, k_1, k_2}
\frac{1}{n\hbar \omega_e} 2\Re\left[
V^{ee}_{nk,0k_1,00,0k_2}V^{eh *}_{nk,0k_1,00,0k_2}
\right].
\end{equation}
The first term is the increase in binding energy of the exciton in the
absence of the filled Landau level.  The second term is a reduction in
the binding energy (since $V^{ee}$ and $V^{eh}$ have opposite signs);
this may be viewed as the screening of the electron-hole interaction
by this filled Landau level.  Depending on the balance of the two
terms, the overall binding energy of the exciton can either increase
or decrease.

Calculating the matrix elements for Coulomb interactions between the
electrons and for the force-law (\ref{eq:veh}) between the electron
and hole and performing the sums over momenta, we find
\begin{equation}
\Delta B_{X} = +\left[
\sum_{n\neq 0} \frac{\left[{\cal I}_{2n}(d/\ell)\right]^2}
{2^{2n}n(n!)^2}\right] \frac{\left(e^2/4\pi\epsilon\epsilon_0\ell\right)^2}
{\hbar(\omega_e+\omega_h)}
- \left[
\sum_{n\neq 0} \frac{1}{n n! 2^{2n-1}}{\cal I}_{2n-1}(d/\sqrt{2}\ell)
\right] \frac{\left(e^2/4\pi\epsilon\epsilon_0\ell\right)^2}
{\hbar\omega_e},
\end{equation}
where we have defined a function
\begin{equation}
\label{eq:ifun}
{\cal I}_m(z)\equiv \int_0^\infty q^m e^{-q^2/2}e^{-qz} dq.
\end{equation}
For the case $d=0$, the numerical summation of the first term has
previously been presented in the context of the two-dimensional
exciton in an empty quantum well\cite{macdonaldritchie}, and the
second term may be summed exactly\cite{gradshteyn}. The result is
\begin{equation}
\Delta B^{(2)}_{X}(d=0) = +0.44010149\frac{\pi}{2}
\frac{\left(e^2/4\pi\epsilon\epsilon_0\ell\right)^2}
{\hbar(\omega_e+\omega_h)} - \left[\frac{\pi^2}{12}-\frac{(\ln
2)^2}{2}\right]
\frac{\left(e^2/4\pi\epsilon\epsilon_0\ell\right)^2}
{\hbar\omega_e}.
\end{equation} 
For the case of non-zero $d$, used in Sec.~\ref{subsec:positive}, we
have computed the sums numerically.

\end{appendix}

\narrowtext

\widetext


\begin{thebibliography}{10}

\bibitem{clarkreview}
R.~G. Clark,  in {\em Low Dimensional Electronic Systems: New Concepts},
  Vol.~111 of {\em Springer Series in Solid-State Sciences}, edited by G.
  Bauer, F. Kuchar, and H. Heinrich (Springer-Verlag, Berlin, Heidelberg,
  1992), pp.\ 239--255.

\bibitem{heimanb}
D. Heiman {\it et~al.}, Physica B {\bf 201},  315  (1994).

\bibitem{turberfieldpol}
A.~J. Turberfield {\it et~al.}, Phys. Rev. B {\bf 47},  4794  (1993).

\bibitem{goldbergoverview}
B.~B. Goldberg {\it et~al.}, Surf. Sci. {\bf 263},  9  (1992).

\bibitem{uenoyamasham}
T. Uenoyama and L.~J. Sham, Phys. Rev. B {\bf 39},  11044  (1989).

\bibitem{bauer}
G.~E.~W. Bauer, Phys. Rev. B {\bf 45},  9153  (1992).

\bibitem{whittaker}
D.~M. Whittaker, R.~J. Elliott, and J.~M. Rorison, Solid State Commun. {\bf
  75},  703  (1990).

\bibitem{apalkovpikusefros}
V.~M. Apalkov, F.~G. Pikus, and E.~I. Rashba, Phys. Rev. B {\bf 52},  6111
  (1995).

\bibitem{apalkovprb}
V.~M. Apalkov and E.~I. Rashba, Phys. Rev. B {\bf 46},  1628  (1992).

\bibitem{chenquinn95}
X.~M. Chen and J.~J. Quinn, Phys. Rev. B {\bf 51},  5578  (1995).

\bibitem{chenprb94}
X.~M. Chen and J.~J. Quinn, Phys. Rev. B {\bf 50},  2354  (1994).

\bibitem{zangbirman95}
J. Zang and J.~L. Birman, Phys. Rev. B {\bf 51},  5574  (1995).

\bibitem{macrezkell}
A.~H. MacDonald, E.~H. Rezayi, and D. Keller, Phys. Rev. Lett. {\bf 68},  1939
  (1992).

\bibitem{tatarinova}
T.~V. Tatarinova, E.~I. Rashba, and A.~L. Efros, Phys. Rev. B {\bf 50},  17349
  (1994).

\bibitem{rashbaanyon}
E.~I. Rashba and M.~E. Portnoi, Phys. Rev. Lett. {\bf 70},  3315  (1993).

\bibitem{cooperwc}
N.~R. Cooper, Phys. Rev. B {\bf 53},  10804  (1996).

\bibitem{buhmannfqhe}
H. Buhmann {\it et~al.}, Phys. Rev. Lett. {\bf 65},  1056  (1990).

\bibitem{kuktriangle}
I.~V. Kukushkin {\it et~al.}, Phys. Rev. Lett. {\bf 72},  3594  (1994).

\bibitem{plentz}
F. Plentz {\it et~al.},  in {\em High Magnetic Fields in the Physics of
  Semiconductors}, edited by D. Heiman (World Scientific, Singapore, 1995),
  pp.\ 320--323.

\bibitem{sondhi}
S.~L. Sondhi, A. Karlhede, S.~A. Kivelson, and E.~H. Rezayi, Phys. Rev. B {\bf
  47},  16419  (1993).

\bibitem{fertigskyrmion}
H.~A. Fertig, L. Brey, R. Cote, and A.~H. MacDonald, Phys. Rev. B {\bf 50},
  11018  (1994).

\bibitem{macdonaldskyrmion}
A. MacDonald, H. Fertig, and L. Brey, Phys. Rev. Lett. {\bf 76},
2163 (1996).

\bibitem{barrett}
S.~E. Barrett {\it et~al.}, Phys. Rev. Lett. {\bf 74},  5112  (1995).

\bibitem{schmeller}
A. Schmeller, J.~P. Eisenstein, L.~N. Pfeiffer, and K.~W. West, Phys. Rev.
  Lett. {\bf 75},  4290  (1995).

\bibitem{aifer}
E.~H. Aifer, B.~B. Goldberg, and D.~A. Broido, Phys. Rev. Lett. {\bf 76},  680
  (1996).

\bibitem{muzykantskii}
B.~A. Muzykantskii, Zh. Eksp. Teor. Fiz. {\bf 101},  1084  (1992), [Sov. Phys.
  {J}{E}{T}{P}{\bf 74}, 897 (1992)].

\bibitem{ekenbergaltarelli}
U. Ekenberg and M. Altarelli, Phys. Rev. B {\bf 32},  3712  (1985).

\bibitem{footnote}
In some single-heterojunction samples higher subband states of the
electron are sometimes
observed\protect\cite{clarkreview,turberfieldpol}. Our theory cannot
be directly applied to~such samples.

\bibitem{prangeandgirvin}
{\em The Quantum {H}all Effect}, 2nd  ed., edited by R.~E. Prange and S.~M.
  Girvin (Springer-Verlag, Berlin, 1990).

\bibitem{bychkovrashba}
Y.~A. Bychkov and E.~I. Rashba, Solid State Commun. {\bf 48},  399  (1983).

\bibitem{paquetriceueda}
D. Paquet, T.~M. Rice, and K. Ueda, Phys. Rev. B {\bf 32},  5208  (1985).

\bibitem{dzyubenkolozovik}
A.~B. Dzyubenko and Y.~E. Lozovik, J. Phys. A {\bf 24},  415  (1991).

\bibitem{vietbirman}
N.~A. Viet and J.~L. Birman, Phys. Rev. B {\bf 51},  14337  (1995).

\bibitem{flavio}
F. Plentz and D. Heiman, private communication.

\bibitem{eisensteinholemass}
J.~P. Eisenstein {\it et~al.}, Phys. Rev. Lett. {\bf 53},  2579  (1984).

\bibitem{wojshawrylak}
A. Wojs and P. Hawrylak, Phys. Rev. B {\bf 51},  10880  (1995).

\bibitem{haldanehierarchy}
F.~D.~M. Haldane, Phys. Rev. Lett. {\bf 51},  605  (1983).

\bibitem{fanoortolani}
G. Fano, F. Ortolani, and E. Colombo, Phys. Rev. B {\bf 34},  2670  (1986).

\bibitem{bychkov}
Y.~A. Bychkov, S.~V. Iordanskii, and G.~M. Eliashberg, Pis'ma Zh. Eksp. Teor.
  Fiz. {\bf 33},  152  (1981), [Sov. Phys. {J}{E}{T}{P}~{\bf 33}, 143 (1981)].

\bibitem{kallinhalperin1}
C. Kallin and B.~I. Halperin, Phys. Rev. B {\bf 30},  5655  (1984).

\bibitem{gorkov}
L.~P. Gor'kov and I.~E. Dzyaloshinskii, Zh. Eksp. Teor. Fiz. {\bf 53},  717
  (1967), [Sov. Phys. {J}{E}{T}{P}~{\bf 26}, 449 (1969)].

\bibitem{lernloz}
I.~V. Lerner and Y.~E. Lozovik, Zh. Eksp. Teor. Fiz. {\bf 78},  1167  (1978),
  [Sov. Phys. {J}{E}{T}{P}~{\bf 51}, 588 (1980)].

\bibitem{kukushkindos}
I.~V. Kukushkin, S.~V. Meshkov, and V.~B. Timofeev, Usp. Fiz. Nauk Sov. Phys.
  Usp. {\bf 155},  219  (1988), [Sov. Phys. Usp.{\bf 31}, 511 (1988)].

\bibitem{pikusefros}
F.~G. Pikus and A.~L. Efros, Sov. Phys. JETP {\bf 69},  558  (1989).

\bibitem{halperinlax1}
B.~I. Halperin and M. Lax, Phys. Rev. {\bf 148},  722  (1966).

\bibitem{halperinlax2}
B.~I. Halperin and M. Lax, Phys. Rev. {\bf 153},  802  (1967).

\bibitem{lifshitz1}
I.~M. Lifshitz, Usp. Fiz. Nauk {\bf 83},  617  (1964), [Sov. Phys. Usp. {\bf
  7}, 549 (1965)].

\bibitem{lifshitz2}
I.~M. Lifshitz, Zh. Eksp. Teor. Fiz. {\bf 53},  743  (1967), [Sov. Phys.
  {J}{E}{T}{P}~{\bf 26}, 462 (1968)].

\bibitem{noneqexciton}
M. Gurioli, A. Vinattieri, J. Martinez-Pastor, and M. Colocci, Phys. Rev. B {\bf
  50},  11817  (1994).

\bibitem{yang}
F. Yang, M. Wilkinson, E.~J. Austin, and K.~P. Odonnell, Phys. Rev. Lett. {\bf
  70},  323  (1993).

\bibitem{qmcgap}
B. Kralik, A.~M. Rappe, and S.~G. Louie, Phys. Rev. B {\bf 52},  11626  (1995).

\bibitem{heimangates}
F. Plentz {\it et~al.},  {\em CMMP 1995 conference proceedings}, pp.~98--101.

\bibitem{rezayi}
E.~H. Rezayi, Phys. Rev. B {\bf 43},  5944  (1991).

\bibitem{dzyubenko}
A.~B. Dzyubenko, Physics Letters {A} {\bf 173},  311  (1993).

\bibitem{wilkinson}
M. Wilkinson, F. Yang, E.~J. Austin, and K.~P. O'Donnell, J. Phys. C {\bf 4},
  8863  (1992).

\bibitem{macdonaldritchie}
A.~H. Macdonald and D.~S. Ritchie, Phys. Rev. B {\bf 33},  8336  (1986).

\bibitem{gradshteyn}
A.~S. Gradshteyn and I.~M. Ryzhik, {\em Tables of Integrals, Series, and
  Products} (Academic Press, New York, 1980).

\end{thebibliography}
\end{document}